\PassOptionsToPackage{table}{xcolor}
\PassOptionsToPackage{hyphens}{url}
\documentclass[sigplan,authorversion]{acmart}

\setcopyright{acmlicensed}
\acmPrice{15.00}
\acmDOI{10.1145/3278122.3278125}
\acmYear{2018}
\copyrightyear{2018}
\acmISBN{978-1-4503-6045-6/18/11}
\acmConference[GPCE '18]{Proceedings of the 17th ACM SIGPLAN International Conference on Generative Programming: Concepts and Experiences}{November 5--6, 2018}{Boston, MA, USA}
\acmBooktitle{Proceedings of the 17th ACM SIGPLAN International Conference on Generative Programming: Concepts and Experiences (GPCE '18), November 5--6, 2018, Boston, MA, USA}

\usepackage[T1]{fontenc}
\usepackage[all]{nowidow}
\usepackage{microtype}
\usepackage{accsupp}
\usepackage{rascal-lst}
\usepackage{listings}
\usepackage{cleveref}
\usepackage{mathtools}
\usepackage{stmaryrd}
\usepackage{amsmath,amsfonts,amssymb}
\usepackage{pifont}
\usepackage{float}
\floatstyle{plaintop}
\restylefloat{table}
\usepackage{thmtools}
\usepackage{thm-restate}
\usepackage{booktabs}
\usepackage{multirow}
\usepackage{multicol}
\usepackage{semantic}
\usepackage{array}
\usepackage{galois}
\usepackage[]{todonotes}
\usepackage{bbold}
\usepackage[inline]{enumitem}
\usepackage{placeins}
\usepackage{tcolorbox}
\usepackage{mdframed}
\usepackage{xpatch}
\usepackage{tikz}
\usepackage{subcaption}
\usepackage{flushend}

\makeatletter
\xpatchcmd{\endmdframed}
  {\aftergroup\endmdf@trivlist\color@endgroup}
  {\endmdf@trivlist\color@endgroup\@doendpe}
  {}{}
  \makeatother

\usepackage[scale=.85,defaultmono]{droidmono}

\newcommand*{\llbrace}{%
  \BeginAccSupp{method=hex,unicode,ActualText=2983}%
    \textnormal{\usefont{OMS}{lmr}{m}{n}\char102}%
    \mathchoice{\mkern-4.05mu}{\mkern-4.05mu}{\mkern-4.3mu}{\mkern-4.8mu}%
    \textnormal{\usefont{OMS}{lmr}{m}{n}\char106}%
  \EndAccSupp{}%
}
\newcommand*{\rrbrace}{%
  \BeginAccSupp{method=hex,unicode,ActualText=2984}%
    \textnormal{\usefont{OMS}{lmr}{m}{n}\char106}%
    \mathchoice{\mkern-4.05mu}{\mkern-4.05mu}{\mkern-4.3mu}{\mkern-4.8mu}%
    \textnormal{\usefont{OMS}{lmr}{m}{n}\char103}%
  \EndAccSupp{}%
}

\newcommand{\Rho}{\scalebox{1.2}{$\varrho$}}

\newcommand{\rulename}[1]{\textsc{\scriptsize{#1}}}
\newcommand{\keyword}[1]{\text{\upshape \textsf{#1}}}

\newcommand{\namedset}[1]{\ensuremath{\mathsf{#1}}}
\newcommand{\anamedset}[1]{\ensuremath{\abst{\namedset{#1}}}}
\newcommand{\abst}[1]{\ensuremath{\widehat{#1}}}
\newcommand{\auxiliary}[1]{\textsf{#1}}
\newcommand{\aauxiliary}[1]{\abst{\textsf{#1}}}
\newcommand{\tuple}[1]{\ensuremath{(#1)}}
\newcommand{\ttuple}[1]{\ensuremath{\langle #1 \rangle}}
\newcommand{\esequence}[1]{\underline{\ensuremath{#1}}}
\newcommand{\sem}[1]{\llbracket#1\rrbracket}
\newcommand{\satsem}[3]{#1 |=_{#2} #3}
\newcommand{\satsemstar}[3]{#1 |=^{\star}_{#2} #3}
\newcommand{\abssatsem}[3]{#1 ~\abst{|=}_{#2}~ #3}
\newcommand{\abssatsemstar}[3]{#1 ~\abst{|=}^{\star}_{#2}~ #3}
\newcommand{\collect}[1]{\llbrace #1 \rrbrace}
\newcommand{\evalexpr}[4]{#1; #2 \xRightarrow[\mathrm{expr}]{} #3; #4}
\newcommand{\match}[4]{#3 |- #1 \overset{?}{\coloneqq} #2 \xRightarrow[\textrm{match}]{} #4}
\newcommand{\matchall}[6]{#3 |- #1 \overset{?}{\coloneqq} #2 ~|~ #4 \xRightarrow[\textrm{match$\star$}]{#5} #6}

\newcommand{\oprev}{\ensuremath{o_\text{\textrm{prev}}}}
\newcommand{\fail}{\ensuremath{\keyword{fail}}}
\newcommand{\success}{\ensuremath{\keyword{success}}}
\newcommand{\Nat}{\ensuremath{\namedset{Nat}}}
\newcommand{\Expr}{\ensuremath{\namedset{Expr}}}
\newcommand{\str}{\ensuremath{\namedset{str}}}
\newcommand{\zero}{\ensuremath{\auxiliary{zero}}}
\newcommand{\suc}[1]{\ensuremath{\auxiliary{suc}\left(#1\right)}}
\newcommand{\cst}[1]{\ensuremath{\auxiliary{cst}\left(#1\right)}}
\newcommand{\mult}[1]{\ensuremath{\auxiliary{mult}\left(#1\right)}}
\newcommand{\var}[1]{\ensuremath{\auxiliary{var}\left(#1\right)}}
\newcommand{\hit}{\textsf{hit:}}

\newcommand{\subtyping}[2]{#1 ~{<:}~ #2}
\newcommand{\atyping}[2]{#1 ~\abst{:}~ #2}
\newcommand{\asubtyping}[2]{#1 ~\abst{<:}~ #2}
\newcommand{\anotsubtyping}[2]{#1 ~\abst{\rlap{\smash{\ensuremath{\not<:}}}\phantom{<:}}~ #2}
\newcommand{\aevalexpr}[3]{#1; #2 \xRightarrow[\textrm{a-expr}]{} #3}
\newcommand{\aevalexprc}[4]{#1; #2 \xRightarrow[\textrm{a-expr-{#4}}]{} #3}
\newcommand{\aevalexprstar}[3]{#1; #2 \xRightarrow[\textrm{a-expr$\star{}$}]{} #3}
\newcommand{\aevalexprstarc}[4]{#1; #2 \xRightarrow[\textrm{a-expr$\star{}$-{#4}}]{} #3}
\newcommand{\amatch}[4]{#3 |- #1 \overset{?}{\coloneqq} #2 \xRightarrow[\textrm{a-match}]{} #4}
\newcommand{\amatchall}[5]{#3 |- #1 \overset{?}{\coloneqq} #2 , #4 \xRightarrow[\textrm{a-match$\star$}]{} #5}
\newcommand{\amatchc}[6]{#3 |- #1 \overset{?}{\coloneqq} #2
  \ifx #4\empty \else ~|~ #4 \fi \xRightarrow[\textrm{a-match-{#6}}]{} #5}
\newcommand{\amatchallc}[6]{#3 |- #1 \overset{?}{\coloneqq} #2
\ifx #4\empty \else , #4 \fi \xRightarrow[\textrm{a-match$\star$-{#6}}]{} #5}
\newcommand{\aevalbuvisit}[4]{#1; #2; #3 \xRightarrow[\mathrm{a-bu-visit}]{} #4}
\newcommand{\aevalbuvisitc}[5]{#1; #2; #3 \xRightarrow[\mathrm{a-bu-visit-{#5}}]{} #4}
\newcommand{\aevalbuvisitstar}[4]{#1; #2; #3 \xRightarrow[\mathrm{a-bu-visit\star{}}]{} #4}
\newcommand{\aevalbuvisitstarc}[5]{#1; #2; #3 \xRightarrow[\mathrm{a-bu-visit\star{}-{#5}}]{} #4}
\newcommand{\areconstruct}[3]{\abst{\textsf{recons }} #1 \textsf{ using } #2 \textsf{ to } #3}
\newcommand{\aevalcases}[4]{#1; #2; #3 \xRightarrow[\mathrm{a-cases}]{} #4}
\newcommand{\aevalcasesc}[5]{#1; #2; #3 \xRightarrow[\mathrm{a-cases-{#5}}]{} #4}
\newcommand{\aevalcase}[4]{#1; #2; #3 \xRightarrow[\mathrm{a-case}]{} #4}
\newcommand{\aevalcasec}[5]{#1; #2; #3 \xRightarrow[\mathrm{a-case-{#5}}]{} #4}
\newcommand{\maybe}[1]{#1^{?}}

\definecolor{psred}{RGB}{215,25,28}
\definecolor{psorange}{RGB}{253,174,97}
\definecolor{psyellow}{RGB}{255,255,191}
\definecolor{psgreen}{RGB}{171,221,164}
\definecolor{psblue}{RGB}{43,131,186}

\crefname{figure}{\text{Fig.}}{\text{figures}}
\Crefname{figure}{\text{Figure}}{\text{Figures}}
\crefname{section}{\text{Sect.}}{\text{sections}}
\Crefname{section}{\text{Section}}{\text{Sections}}
\crefformat{section}{Sect.\,#2#1#3}
\Crefformat{section}{Section\,#2#1#3}
\crefformat{figure}{Fig.\,#2#1#3}
\Crefformat{figure}{Figure\,#2#1#3}

\makeatletter
\newcommand\incircbin
{%
  \mathpalette\@incircbin
}
\newcommand\@incircbin[2]
{%
  \mathbin%
  {%
    \ooalign{\hidewidth$#1#2$\hidewidth\crcr$#1\bigcirc$}%
  }%
}

\makeatother

\newcommand{\cmark}{\ding{51}}%
\newcommand{\xmark}{\ding{55}}%

\begin{document}
%
\title[Verification of High-Level Transformations ...]{Verification of High-Level Transformations with Inductive Refinement Types}



\author{Ahmad Salim Al-Sibahi}

\affiliation{
  \institution{IT University of Copenhagen}
}
\affiliation{
  \institution{University of Copenhagen}
}

\affiliation{
  \institution{Skanned.com}            
  \country{Denmark}                    
}

\email{ahmad@{di.ku.dk, skanned.com}}          

\author{Thomas P. Jensen}
\affiliation{
  \institution{Inria Rennes}           
  \country{France}                   
}
\email{thomas.jensen@inria.fr}         

\author{Aleksandar S. Dimovski}
\affiliation{
  \institution{IT University of Copenhagen}
  \country{Denmark}
}
\affiliation{
  \institution{Mother Teresa University, Skopje}            
  \country{Macedonia}                    
}
\email{aleksandar.dimovski@unt.edu.mk}          

\author{Andrzej W\k{a}sowski}
\affiliation{
  \institution{IT University of Copenhagen}            
  \country{Denmark}                    
}
\email{wasowski@itu.dk}          

\begin{abstract}
High-level transformation languages like Rascal include expressive features for manipulating
large abstract syntax trees: first-class traversals, expressive
pattern matching, backtracking and generalized iterators. We present
the design and implementation of an abstract
interpretation tool, Rabit, for verifying inductive type and shape properties for
transformations written in such languages. We describe how to perform abstract
interpretation based on operational semantics,
specifically focusing on the challenges arising when analyzing the expressive
traversals and pattern matching. Finally, we evaluate Rabit on a series of
transformations (normalization, desugaring, refactoring, code
generators, type inference, etc.) showing that we can effectively verify stated properties.
\end{abstract}



\maketitle

\renewcommand{\shortauthors}{A. S. Al-Sibahi, T. P. Jensen, A. S. Dimovski and A.
  W\k{a}sowski}
\section{Introduction}\label{sec:introduction}

Transformations play a central role in software development. They are used, amongst others, for desugaring, model
transformations, refactoring, and code generation. The artifacts involved in
transformations---e.g., structured data, domain-specific models, and
code---often have large abstract syntax, spanning hundreds of syntactic
elements, and a correspondingly rich semantics.  Thus, writing transformations
is a tedious and error-prone process.
Specialized languages and frameworks with high-level features have been developed to address
this challenge of writing and
maintaining transformations.
These languages include Rascal\,\cite{Klint2011}, Stratego/XT\,\cite{DBLP:journals/scp/BravenboerKVV08},
TXL\,\cite{DBLP:journals/scp/Cordy06},
Uniplate\,\cite{DBLP:conf/haskell/MitchellR07} for Haskell, and
Kiama\,\cite{SPRINGER:conf/gtse/Sloane11} for Scala.
For example, Rascal 
combines a functional
core language supporting state and exceptions, with constructs for processing of large structures.
\begin{figure}[ht]
  \centering
  \begin{minipage}{1.0\linewidth}
  \begin{lstlisting}[language=Rascal,xleftmargin=-2pt]
public Script flattenBlocks(Script s) {
  solve(s) {
    s = bottom-up visit(s) {
      case stmtList: [*xs,block(ys),*zs] =>
                xs + ys + zs
    }
  }
  return s;
}
\end{lstlisting}
  \end{minipage}
  \caption{Transformation in Rascal that flattens all nested blocks in a
    statement}
  \label{fig:phpflatten}
\end{figure}

\Cref{fig:phpflatten} shows an example Rascal transformation program taken from a PHP
analyzer.\footnote{\url{https://github.com/cwi-swat/php-analysis}} This transformation
program recursively flattens all blocks in a list of statements. The program
uses the following core Rascal features:
\begin{itemize}

  \item A \emph{visitor} (\lstinline[language=Rascal]{visit}) to traverse and rewrite all statement lists containing a
    block to a flat list of statements. Visitors support various
    strategies, like the \lstinline[language=Rascal]{bottom-up} strategy that traverses the abstract syntax tree starting
    from leaves toward the root.

  \item An \emph{expressive pattern matching} language is used to non-deterministically find
    blocks inside a list of statements. The starred variable
    patterns \lstinline{*xs} and \lstinline{*zs} match arbitrary
    number of elements in the list, respectively before and after the \lstinline{block(ys)} element.
    Rascal supports non-linear matching, negative matching and specifying
    patterns that match deeply nested values.

  \item The solve-loop (\lstinline{solve}) performing the rewrite until a fixed point is
    reached  (the value of $s$ stops changing).
\end{itemize}

\noindent
To rule out errors in transformations, we propose a static analysis
for enforcing type
and shape properties, so that target transformations produce output adhering to
particular shape constraints. For our PHP example, this would include:
\begin{itemize}
\item The transformation preserves the constructors used in the input: does
  not add or remove new types of PHP statements.
\item The transformation produces flat statement lists, i.e., lists
  that do not recursively contain any block.
\end{itemize}
To ensure such properties, a verification technique must reason about shapes of
inductive data---also inside collections such as sets and maps---while
still maintaining soundness and precision. It must also track other
important aspects, like cardinality of collections, which interact
with target language operations including pattern matching and
iteration.

In this paper, we address the problem of verifying type and shape
properties for high-level transformations written in Rascal and
similar languages.  We show how to design and implement a static
analysis based on abstract interpretation.  Concretely, our
contributions are:
\begin{enumerate}

  \item An abstract interpretation-based static analyzer---Rascal ABstract
    Interpretation Tool (Rabit)---that supports inferring types and inductive
    shapes for a large subset of Rascal.

  \item An evaluation of Rabit on several program transformations: refactoring,
    desugaring, normalization algorithm, code generator, and language
    implementation of an expression language.

  \item A modular design for abstract shape domains, that allows extending and replacing abstractions for concrete element types, e.g.  extending the abstraction for lists to include length in addition to shape of contents.

  \item Schmidt-style abstract \emph{operational}
    semantics\,\cite{DBLP:journals/lisp/Schmidt98} for a significant subset of
    Rascal adapting the idea of \emph{trace memoization} to support arbitrary recursive calls with input from infinite domains.

\end{enumerate}
Together, these contributions show feasibility of applying abstract
interpretation for constructing analyses for expressive transformation languages and properties.

We proceed by presenting a running example in \cref{sec:motivationandidea}.
We introduce the key constructs of Rascal in
\cref{sec:formallanguage}. \Cref{sec:abstractdomains} describes the modular
construction of abstract domains.
\Crefrange{sec:abstractsemantics}{sec:memoizationstrategies}
describe abstract semantics. We evaluate the analyzer
on realistic transformations, reporting results in \cref{sec:evaluation}. \Cref{sec:relatedwork,sec:conclusion} discuss related papers and conclude.


\section{Motivation and Overview}\label{sec:motivationandidea}
Verifying types and state properties such as the ones stated for the
program of \cref{fig:phpflatten} poses the following key challenges:
\begin{itemize}

  \item The programs use \emph{heterogeneous inductive data types}, and contain \emph{collections} such as lists, maps and sets, and basic data such as integers and strings. This complicates construction of the abstract domains, since one shall model interaction between these different types while maintaining precision.

  \item The traversal of syntax trees depends heavily on the \emph{type and shape of input}, on a \emph{complex program state}, and involves \emph{unbounded recursion}. This challenges the inference of approximate invariants in a procedure that both terminates and provides useful results.

  \item Backtracking and exceptions in large programs introduce the possibility
    of \emph{state-dependent non-local jumps}. This makes it
    difficult to statically calculate the control
    flow of target programs and have a compositional denotational semantics,
    instead of an operational one.

\end{itemize}

\begin{figure}[t] 
  \centering
  \begin{minipage}{1.0\linewidth}
  \begin{lstlisting}[xleftmargin=-4pt]
 data Nat = zero() | suc(Nat pred);
 data Expr = var(str nm) | cst(Nat vl)
           | mult(Expr el, Expr er);

 Expr simplify(Expr expr) =
   bottom-up visit (expr) {
       case mult(cst(zero()), y) => cst(zero())
       case mult(x, cst(zero())) => cst(zero())
   };
  \end{lstlisting}
  \end{minipage}
  \caption{The running example: eliminating multiplications by \zero\ from expressions}
  \label[figure]{fig:runningexample}
\end{figure} 

\noindent\Cref{fig:runningexample} presents a small pedagogical example using
visitors. The program performs expression simplification by
 traversing a syntax tree bottom-up and
reducing multiplications by constant zero.
We now survey the analysis techniques contributed in this
paper, explaining them using this example.

\paragraph{Inductive refinement types}
Rabit works by inferring an inductive refinement type representing the shape of
possible output of a transformation given the shape of its input. It does this by interpreting
the simplification program abstractly, considering all possible paths the
program can take for values satisfying the input shape (any
expression of type $\namedset{Expr}$ in this case). The result of running Rabit on this case is:
\begin{align*}
& \success\, \cst\Nat \wr \var\str \wr \mult{\Expr',\Expr'}\\
& \fail\, \cst\Nat \wr \var\str \wr \mult{\Expr',\Expr'}
\end{align*}
where
$\Expr' = \cst{\suc\Nat} \wr \var\str \wr \mult{\Expr', \Expr'}$.

We briefly interpret how to read this type. The bar $\wr$ denotes a choice between alternative constructors.  If the input was rewritten during traversal (\success, the first line) then
the resulting syntax tree contains no multiplications by \zero. All multiplications may only involve $\Expr'$, which disallows the zero constant at the top level. Observe this in the last alternative $\mult{\Expr',\Expr'}$ that contains only
expressions of type $\Expr'$, which in turn only allows
multiplications by constants constructed using $\suc{\Nat}$ (that is
$\geq 1$). If
the traversal failed to match (\fail, the second line), then the input did not
contain any multiplication by \zero{} to begin with and so does not the output, which has not been rewritten.

The success and failure happen to be the same for our example, but
this is not necessarily always the case. Keeping separate result values allows
retaining precision throughout the traversal, better reflecting concrete execution paths.
We now proceed discussing how Rabit can infer this shape using abstract interpretation.

\newcommand{\memo}[4]{{%
  \scriptsize
  \renewcommand{\arraystretch}{1.1}
  \rowcolors{1}{psorange}{psorange}
  \begin{tabular}{|ll|}\hline
    \namedset{\strut Expr}\hspace{-1.4mm} & $\strut \mapsto\! #1,#2$\\
    \namedset{\strut Nat}\hspace{-1.4mm}  & $\strut \mapsto\! #3,#4$\\\hline
  \end{tabular}%
}}

\newcommand{\nemo}[2]{{
  \scriptsize
  \renewcommand{\arraystretch}{1.1}
  \rowcolors{1}{psorange}{psorange}
  \begin{tabular}{|ll|}\hline
    \namedset{\strut Nat}\hspace{-1.4mm} & $\strut \mapsto\! #1,#2$\\\hline
  \end{tabular}%
}}

\newcommand{\labelnode}[2]{%
  \node[circle,minimum height=1em,fill=psblue,draw=gray,text=psyellow,ultra thin,inner sep=.1mm] at (#1.south east)  {\footnotesize\textbf{\textsf{#2}}}}

\newcommand{\nodebody}[2]{\strut\,\textsf{input:}\,\(#1\)\\[.5mm] #2}
\newcommand{\partition}[1]{\(#1\)}
\newcommand{\widenrec}[4]{\strut \textsf{\,input:}\,\(#1\)\\[-.0mm] #2\\[.4mm]\strut\textsf{\,widen:\,}\(#3\)\\[.5mm]\(\,\oprev=#4\)}

\newcommand{\hs}{25mm}

\tikzset{
  node distance=11mm,
  state/.style={fill=psyellow!70,aetw,inner sep=.7mm,ultra thin,draw=gray},
  aetw/.style={text width=8.0mm},
  aetwr/.style={text width=11.0mm},
  aetwrr/.style={text width=15.9mm},
  aetwrrr/.style={text width=20.5mm},
  aetwrrrr/.style={text width=25.5mm},
  aetwrrrrr/.style={text width=31mm},
  aetwrrrrrr/.style={text width=36.5mm},
  aetwrrrrrrr/.style={text width=41mm},
  aetwrrrrrrrr/.style={text width=47mm},
  aetwrrrrrrrrr/.style={text width=53mm},
  aetwrrrrrrrrrr/.style={text width=60mm},
  partn/.style={fill=psblue!50}
}

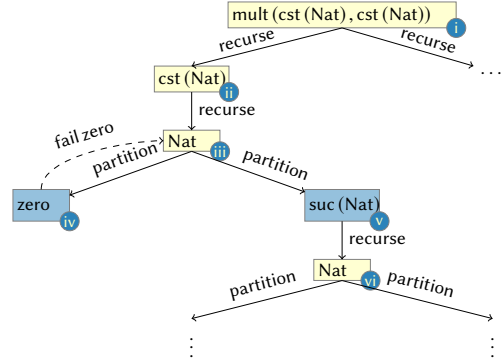
\begin{figure}[t] 
  \hspace{-4mm}\scalebox{.80}{\begin{tikzpicture}
  \small\sffamily

  \node[state,aetwrrrrrr] (root)
  {\mult{\cst{\Nat}, \cst{\Nat}}};
  \labelnode{root}{i};

  \node[state,aetwr] (l) [below=2em of root,xshift=-\hs]
  {\cst{\Nat}};
  \labelnode{l}{ii};

  \draw[->](root.south) to node[sloped,above,xshift=-3mm] {recurse} (l.north);

  \node(r) [below=2em of root, xshift=\hs] {$\cdots$};

  \draw[->](root.south) to node[above,sloped,xshift=3mm] {recurse} (r.north west);

  \node[state,aetw](ld) [below=2em of l]
  {\Nat};
  \labelnode{ld}{iii};

  \draw[->] (l.south) -- node[right]{recurse} (ld);

  \node[state,partn] (ldl) [below=2em of ld,xshift=-\hs]
  {\partition{\strut\zero}};
  \labelnode{ldl}{iv};

  \node[state,aetwr,partn] (ldr) [below=2em of ld, xshift=\hs]
  {\partition{\strut\suc{\Nat}}};
  \labelnode{ldr}{v};

 \draw[->] (ld.south) -- node[above,sloped] {partition} (ldl);
 \draw[->] (ld.south) -- node[above,sloped,xshift=4mm] {partition} (ldr);

 \draw[->,dashed] (ldl.north) to[out=90,in=180,looseness=.66] node[sloped,above,yshift=1mm] {\fail\,\zero} (ld.west);

  \node[state] (ldrd) [below=2em of ldr]
  {\Nat};
  \labelnode{ldrd}{vi};

  \draw[->] (ldr.south) -- node[right]{recurse} (ldrd.north);

  \node (ldrdl) [below=2em of ldrd,xshift=-\hs]
  {\vdots};

    \node (ldrdr) [below=2em of ldrd, xshift=\hs]
  {\vdots};

   \draw[->] (ldrd.south) -- node[above,sloped]{partition} (ldrdl.north);
   \draw[->] (ldrd.south) -- node[above,sloped]{partition} (ldrdr.north);

\end{tikzpicture}}

\caption{Naively abstractly interpreting the simplification example from
  \cref{fig:runningexample} with initial input $\mult{\cst{\Nat}, \cst{\Nat}}$.
  The procedure does not terminate because of infinite recursion on $\Nat$.}
\label{fig:naivetraversal}
\end{figure}

\noindent\paragraph{Abstractly interpreting traversals} The core idea
of abstractly executing a traversal is similar to concrete execution:
we recursively traverse the input structure and rewrite the values
that match target patterns.  However, because of abstraction we must
make sure to take into account all applicable
paths. \Cref{fig:naivetraversal} shows the execution tree of the
traversal on the simplification example (\cref{fig:runningexample})
when it starts with shape $\mult{\cst{\Nat}, \cst{\Nat}}$. Since there is only one
constructor, it will initially \emph{recurse} down to traverse the
contained values (children) creating a new recursion node (yellow, light shaded) in the
figure (ii) containing the left child $\cst{\Nat}$, and then recurse
again to create a node (iii) containing $\Nat$.  Observe here that
$\Nat$ is an abstract type with two possible constructors ($\zero$,
$\suc{\cdot}$), and it is unknown at time of abstract interpretation,
which of these constructors we have.  When Rabit hits a type or a
choice between alternative constructors, it explores each alternative
separately creating new \emph{partition} nodes (blue, darker). In our example we
partition the $\Nat$ type into its constructors $\zero$ (node iv) and
$\suc{\Nat}$ (node v). The $\zero$ case now represents the first case
without children and we can run the visitor operations on it. Since no
pattern matches $\zero$ it will return a $\fail\,\zero$ result
indicating that it has not been rewritten. For the $\suc{\Nat}$ case
it will try to recurse down to $\Nat$ (node vi) which is equal to
(node iii). Here, we observe a problem: if we continue our traversal
algorithm as is, we will not terminate and get a result.  To provide a
terminating algorithm we will resort to using \emph{trace
  memoization}.

\renewcommand{\hs}{5mm}

\paragraph{Partition-driven trace memoization}
The idea  is to detect the paths where execution recursively meets similar input, merging the new recursive node with the similar previous one, thus creating a loop in the execution tree\,\cite{DBLP:journals/lisp/Schmidt98,DBLP:journals/corr/Rosendahl13}. This loop is then resolved by a fixed-point iteration.

In Rabit, we propose \emph{partition-driven trace
  memoization}, which works with potentially unbounded input like
the inductive type refinements that are supported by our abstraction.
We detect cycles by maintaining a \emph{memoization map} which for each
type---used for partitioning---stores the last traversed value (input) and the last result produced for this value (output).
This memoization map is initialized to map all types to the bottom element ($\bot$)
for both input and output. The evaluation is modified to use the
memoization map, so it checks on each iteration the input $i$ against the map:
\begin{itemize}
\item If the last processed refinement type representing the input $i'$ is
  greater than the current input ($i' \sqsupseteq i$), then it uses the
  corresponding output; i.e., we found a hit in the memoization map.
\item Otherwise, it will merge the last processed and current input refinement types to a new
  value $i'' = i' \nabla i$, update the memoization map and continue execution
  with $i''$. The operation $\nabla$ is called a \emph{widening}; it ensures
  that the result is an upper bound of its inputs, i.e., $i' \sqsubseteq i''
  \sqsupseteq i$ and that the merging will eventually terminate for the
  increasing chain of values. The memoization map is updated to map the general type
  of $i''$ (not refined, for instance \Nat) to map to a pair $\tuple{i'', o}$, where the first component
  denotes the new input $i''$ refinement type and the second component
  denotes the corresponding output $o$ refinement type; initially, $o$ is set to $\bot$ and then
  changed to the result of executing input $i''$ repeatedly until a fixed-point is reached.
\end{itemize}

\begin{figure*}[t] 
\hspace{-1mm}\scalebox{.81}{\begin{tikzpicture}
  \small\sffamily

  \node[state,aetwrrrrr] (root)
  {\widenrec
    {\Nat}
    {\nemo{\bot}{\bot}}
    {\bot\nabla\Nat=\Nat}
    {\bot}};
  \labelnode{root}{1};

  \node[state,partn] (l) [below=of root.south west,anchor=north west] {\partition{\zero}};
  \labelnode{l}{2};

  \node[state,aetwr,partn] (r) [below=of root.south east, anchor=north east]
  {\partition{\suc{\namedset{Nat}}}};
  \labelnode{r}{3};

  \draw[->](root.south) to node[sloped,below,xshift=0mm] {part.} (l.north);
  \draw[->](root.south) to node[above,sloped,xshift=2mm] {part.} (r.north);

  \draw[->,dashed] (l.west) to[out=180,in=180,looseness=.66] node[above,rotate=90,yshift=1mm] {\fail\,\zero} (root.west);

  \node[state,aetwrrr] (rd) [below=of r]
  {\nodebody
      {\Nat}
      {\nemo{\Nat}{\bot}}};
  \labelnode{rd}{4};

  \draw[->] (r) to node[right] {recurse} (rd);

  \draw[->,dashed] (rd.west) to[in=180,out=180,looseness=.66] node[right]{\hit\,$\bot$} (r.west);

  \draw[->,dashed] (r.east) to [out=0,in=0,looseness=.66] node[rotate=90,below] {no reconstruction: $\bot$} (root.east);

  \draw[<-] (root.north)  to node[left] {recurse} ++(up:4mm);



 \node[state,aetwrrrrr] (root2) [right=21mm of root]
  {\widenrec
    {\Nat}
    {\nemo{\bot}{\bot}}
    {\bot\nabla\Nat = \Nat}
    {\fail\,\zero}};
  \labelnode{root2}{1};

  \draw[->,dotted] ([xshift=5mm]root.north)  to[out=90,in=90,looseness=.2] node[above] {\parbox{40mm}{\textsf{output:}\\$\fail\, \bot \nabla (\zero \!\sqcup\!\bot) = \fail\, \zero$}} (root2.north);

  \node[state,partn] (l2) [below=of root2.south west, anchor=north west]
  {\partition{\zero}};
  \labelnode{l2}{2};

  \node[state, aetwr,partn] (r2) [below=of root2.south east, anchor=north east]
  {\partition{\suc{\Nat}}};
  \labelnode{r2}{3};

  \draw[->](root2.south) to node[sloped,below,xshift=1mm] {part.} (l2.north);
  \draw[->](root2.south) to node[above,sloped,xshift=2mm] {partition} (r2.north);

  \draw[->,dashed] (l2.west) to[out=180,in=180,looseness=.66] node[rotate=90,above,yshift=1mm] {\fail\,\zero} (root2.west);

  \node[state,aetwrrrrr] (rd2) [below=of r2]
  {\nodebody
      {\Nat}
      {\nemo{\Nat}{\fail\,\zero}}};
  \labelnode{rd2}{4};

  \draw[->] (r2) to node[right] {recurse} (rd2);

  \draw[->,dashed] (rd2.west) to[in=210,out=180,looseness=.66] node[left]{\parbox{10mm}{\hit\\\fail\,\zero}} (r2.west);

  \draw[->,dashed] (r2.east) to [out=0,in=0,looseness=.66] node[rotate=90,below] {\parbox{5mm}{reconstruction: \fail\,\suc{\zero}}} (root2.east);



 \node[state,aetwrrrrrr] (root3) [right=25mm of root2]
  {\widenrec
    {\Nat}
    {\nemo{\bot}{\bot}}
    {\bot\nabla\Nat = \Nat}
    {\fail\,\zero\wr\suc{\zero}}};
  \labelnode{root3}{1};

  \draw[->,dotted] ([xshift=5mm]root2.north)  to[out=90,in=90,looseness=.2] node[above,xshift=5mm] {\parbox{68mm}{\textsf{output:}\\\(\fail\, \zero \nabla (\zero\!\sqcup\!\suc{\zero})=\fail\, \zero\wr\suc{\zero}\)}} (root3.north);

  \node[state,partn] (l3) [below=of root3.south west, anchor=north west]
  {\partition{\zero}};

  \node[state, aetwr,partn] (r3) [below=of root3.south east, anchor=north east]
  {\partition{\suc{\Nat}}};
  \labelnode{r3}{3};

  \draw[->](root3.south) to node[sloped,below,xshift=1mm] {part.} (l3.north);
  \draw[->](root3.south) to node[above,sloped,xshift=2mm] {part.} (r3.north);

  \draw[->,dashed] (l3.west) to[out=180,in=180,looseness=.66] node[rotate=90,above,yshift=1.3mm] {\fail\,\zero} (root3.west);

  \node[state,aetwrrrrrrr] (rd3) [below=of r3,xshift=-14mm]
  {\nodebody
      {\Nat}
      {\nemo{\Nat}{\fail\,\zero\!\wr\!\suc{\zero}}}};
  \labelnode{rd3}{4};

  \draw[->] (r3) to node[right] {recurse} (rd3);

  \draw[->,dashed] (rd3.west) to[in=180,out=180, looseness=1] node[above,xshift=-8mm]{\parbox{10mm}{\hit\fail\,\zero\\$\wr\,$\suc{\zero}}} (r3.west);

  \draw[->,dashed] (r3.east) to [out=0,in=0, looseness=.4] node[rotate=90,below,xshift=-15mm,yshift=1mm] {\parbox{5mm}{reconstruction: \fail\,\suc{\zero\!\wr\!\suc{\zero}}}} (root3.east);

  \labelnode{l3}{2};

  \node[anchor=west] (stop) at ([xshift=15mm,yshift=3mm]root3.north) {$\cdots \fail\,\Nat$};

  \draw[->,dotted] (root3.north) to[out=90,in=180,looseness=.60] (stop);


\end{tikzpicture}}

\caption{Three iterations of a fixed point computation for input \Nat. Iterations are separated by dotted arrows on top}
\label{fig:nat-iteration}
\end{figure*}
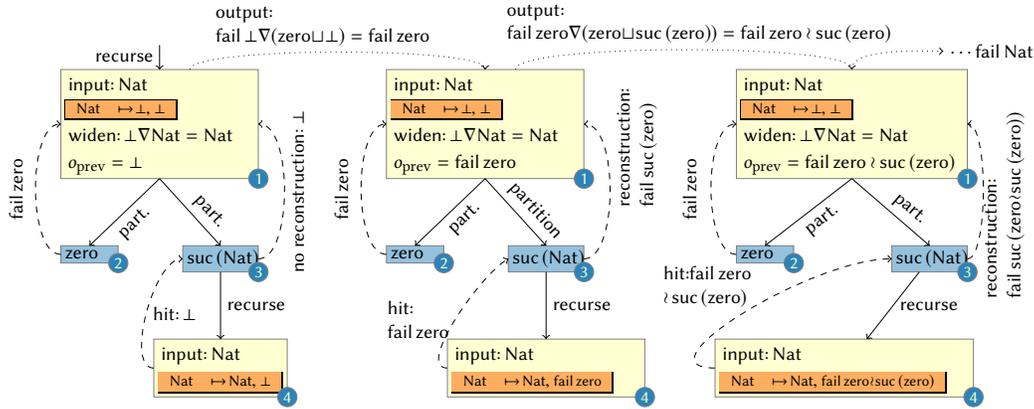

\noindent{}We demonstrate the trace memoization and fixed-point iteration
procedures on \Nat{} in \cref{fig:nat-iteration}, beginning with the
leftmost tree.  The expected result is \fail\,\Nat, meaning that no pattern has
matched, no rewrite has happened, and a value of type \Nat\ is returned, since
the simplification program only introduces changes to values of type \Expr.

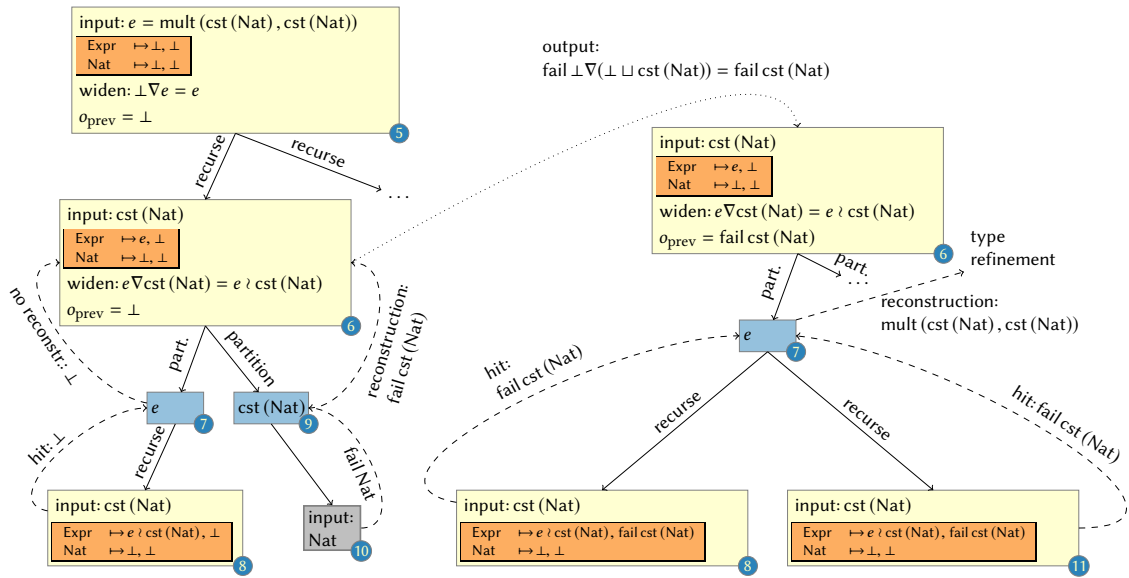
\begin{figure*}[t] 
  \hspace{-4mm}\scalebox{.80}{\begin{tikzpicture}
  \small\sffamily


  \node[state,aetwrrrrrrrrr] (root)
  {\widenrec
    {e=\mult{\cst{\Nat}, \cst{\Nat}}}%
    {\memo{\bot}{\bot}{\bot}{\bot}}
    {\bot\nabla e = e}
    {\bot}};
  \labelnode{root}{5};

  \node[state,aetwrrrrrrrr] (l) [below=of root,xshift=-\hs]
  {\widenrec
     {\cst{\Nat}}%
     {\memo{e}{\bot}{\bot}{\bot}}
     {e \nabla \cst{\Nat} = e \wr \cst{\Nat}}
     {\bot}};
  \labelnode{l}{6};

  \draw[->](root.south) to node[sloped,above,xshift=0mm] {recurse} (l.north);

  \node(r) [right=\hs of l.north east] {$\cdots$};

  \draw[->](root.south) to node[above,sloped,xshift=1mm] {recurse} (r.north west);

  \node[state,partn] (ll) [below=of l,xshift=-\hs]
  {\partition{\strut e}};
  \labelnode{ll}{7};

  \node[state,aetwr,partn] (lr) [right=\hs of ll]
  {\partition{\strut\cst{\Nat}}};
  \labelnode{lr}{9};

 \draw[->] (l.south) -- node[above,sloped] {part.} (ll);
 \draw[->] (l.south) -- node[above,sloped,xshift=2mm] {partition} (lr);

  \node[state,aetwrrrrr] (lll) [below=of ll,xshift=-\hs]
  {\nodebody
      {\cst{\Nat}}
      {\memo{e\wr \cst{\Nat}}{\bot}{\bot}{\bot}}};
  \labelnode{lll}{8};


  \draw[->] (ll.south) -- node[above,sloped] {recurse} (lll.north);

  \draw[->,dashed] (lll) to[out=170,in=180] node[above,sloped]  {\hit\,$\bot$}  (ll);
  \draw[->,dashed] (ll) to[out=170,in=180] node[below,sloped]  {no reconstr.: $\bot$}  (l);

  \node[state, draw=gray,thick,fill=gray!50] (lrd) [right=2.0*\hs of lll]
  {input: \Nat};
  \labelnode{lrd}{10};

  \draw[->] (lr.south) -- (lrd.north);
  \draw[->,dashed] (lrd.east) to[out=0,in=0] node[below,sloped] {\fail\,\Nat}(lr.east);
  \draw[->,dashed] (lr.east) to[out=0,in=0] node[below,sloped] {\parbox{20mm}{reconstruction:\\\fail\,\cst{\Nat}}} (l.east);



  \node[state,aetwrrrrrrrr] (l2) [right=10*\hs of l, yshift=12mm]
  {\widenrec
     {\cst{\Nat}}%
     {\memo{e}{\bot}{\bot}{\bot}}
     {e \nabla \cst{\Nat} = e \wr \cst{\Nat}}
     {\fail\,\cst{\Nat}}};
  \labelnode{l2}{6};

  \draw[->,dotted] (l.east) to[out=30,in=100,looseness=.6] node[above,xshift=15mm,yshift=8mm] {\parbox{50mm}{\textsf{output:}\\$\fail\,\bot \nabla (\bot \sqcup \cst{\Nat})=\fail\,\cst{\Nat}$}} (l2.north);

  \node[state,partn] (ll2) [below=of l2,xshift=-\hs]
  {\partition{\strut e}};
  \labelnode{ll2}{7};

  \node (lr2) at ([shift={(-27:8mm)}]l2.south) [anchor=north west] {$\cdots$};

  \draw[->] (l2.south) -- node[above,sloped] {part.} (ll2);
  \draw[->] (l2.south) -- node[above,sloped,xshift=5mm] {part.} (lr2);

  \node[state,aetwrrrrrrrr] (lll2) [below=23mm of ll2,xshift=-5.5*\hs]
  {\nodebody
      {\cst{\Nat}}
      {\memo{e\wr \cst{\Nat}}{\fail\,\cst{\Nat}}{\bot}{\bot}}};
  \labelnode{lll2}{8};

  \node[state,aetwrrrrrrrr](llr2) [below=23mm of ll2,xshift=5.5*\hs]
  {\nodebody{\cst\Nat}
    {\memo{e\wr\cst{\Nat}}{\fail\,\cst{\Nat}}{\bot}{\bot}}
  };
  \labelnode{llr2}{11};

  \draw[->] (ll2.south) -- node[above,sloped] {recurse} (lll2.north);
  \draw[->] (ll2.south) -- node[above,sloped,xshift=2mm] {recurse} (llr2.north);

  \draw[->,dashed] (lll2) to[out=170,in=180] node[above,sloped,xshift=9mm]  {\parbox{17mm}{\hit\\$\fail\,\cst{\Nat}$}}  (ll2.west);

  \draw[->,dashed] (llr2.east) to[out=0,in=-6,looseness=1.2] node[above,sloped] {\hit\,\fail\,\cst\Nat} (ll2.east);

  \draw[->,dashed] (ll2.north east) to node[right,xshift=-1mm,yshift=-3.0mm]  {\parbox{40mm}{reconstruction:\\\mult{\cst\Nat,\cst\Nat}}} ++(16:29mm) coordinate(aw);

  \node[anchor=south west] at (aw) {\parbox{10mm}{\sffamily type\\refinement}};


\end{tikzpicture}}

\caption{A prefix of the abstract interpreter run for $e=\mult{\cst{\Nat},\cst{\Nat}}$. Fragments of two iterations involving node 6 are shown, separated by a dotted arrow.}
\label{fig:mult-iter}
\end{figure*}

We show the memoization map inside a framed orange box.
The result of the widening is presented below the memoization map. In all cases the widening in \cref{fig:nat-iteration} is trivial, as
it happens against $\bot$.  The final line in node 1  stores the value \oprev\
produced by the previous iteration of the traversal, to establish whether a
fixed point has been reached ($\bot$ initially).

\paragraph{Trace
  partitioning} We
\emph{partition}\,\cite{DBLP:journals/toplas/RivalM07} the abstract value \Nat\ along its constructors: \zero\
and \suc{\cdot} (\cref{fig:nat-iteration}). This partitioning is key to maintain
precision during the abstract interpretation.
As in \cref{fig:naivetraversal}, the left branch fails immediately, since no pattern in
\cref{fig:runningexample} matches \zero.  The right branch descends into a
new recursion over \Nat, with an updated memoization table.  This run
terminates, due to a hit in the memoization map, returning $\bot$.  After returning,
the value of \suc\Nat\ should be reconstructed with the result of traversing the
child \Nat, but since the
result is $\bot$ there is no value to reconstruct with, so $\bot$ is just propagated
upwards. At the return to the last widening node, the values are joined, and
widen the previous iteration result \oprev\,(the dotted arrow on top).  This
process repeats in the second and third iterations, but now the  reconstruction
in node 3 succeeds: the child \Nat\ is replaced by \zero\ and $\keyword{fail} \;
\suc{\zero}$ is returned (dashed arrow from 3 to 1). In the third iteration, we join and widen the following
components (cf.  \oprev\ and the dashed arrows incoming into node 1 in the rightmost column):
\[
  \left[\zero\wr\suc\zero ~\nabla~ \left( \zero \sqcup \suc{\zero\!\wr\!\suc\zero} \right)\right] = \Nat
  \]
Here, the used widening operator\,\cite{DBLP:conf/fpca/CousotC95} accelerates
the convergence by increasing the value to represent the entire type \Nat.  It
is easy to convince yourself, by following the same recursion steps as in the
figure, that the next iteration, using $\oprev=\Nat$ will produce \Nat\ again,
arriving at a fixed point. Observe, how consulting the
memoization map, and widening the current value accordingly, allowed us to avoid infinite recursion over unfoldings of \Nat.

\paragraph{Nesting fixed point iterations.}
When inductive shapes (e.g., \Expr) refer to other inductive shapes
(e.g., \Nat), it is necessary to run nested fixed-point iterations
to solve recursion at each level.
\Cref{fig:mult-iter} returns to the more
high-level fragment of the traversal of \Expr\ starting with
\mult{\cst\Nat,\cst\Nat} as in \cref{fig:naivetraversal}.  We follow the recursion tree along nodes $5, 6, 7, 8$, $9, 10, 9, 6$ with the same rules as in \cref{fig:nat-iteration}.  In node 10 we run a nested fixed point iteration on \Nat, already discussed in \cref{fig:nat-iteration}, so we just include the final result.

\paragraph{Type refinement.} The output of the first iteration in node 6 is \fail\,\cst\Nat, which becomes the new \oprev, and the second iteration begins (to the right). After the widening the input is partitioned into $e$ (node 7) and \cst\Nat (node elided). When the second iteration returns to node 7 we have the following reconstructed value: \mult{\cst\Nat,\cst\Nat}.  Contrast this with lines 6-7 in \cref{fig:runningexample}, to see that running the abstract value against this pattern might actually produce \success.  In order to obtain precise result shapes, we refine the input values when they fail to match a pattern.  Our abstract interpreter produces a refinement of the type, by running it through the pattern matching, giving:
\begin{align*}
  & \success~ \cst\Nat\\
  & \fail~ \mult{
    \cst{\suc\Nat},~
    \cst{\suc\Nat}}
\end{align*}
The result means, that if the pattern match succeeds then it produces an expression of type \cst{\Nat}. More interestingly, if the matching failed neither the left nor the right argument of \mult{\cdot,\cdot}\ could have contained the constant \zero---the interpreter captured some aspect of the semantics of the program by \emph{refining} the input type.
Naturally, from this point on the recursion and iteration continues, but we shall abandon the example, and move on to formal developments.


\section{Formal Language}\label{sec:formallanguage}

The presented technique is meant to be general and applicable to many  high-level
transformation languages. However, to keep the presentation concise, we
focus on few key constructs from Rascal\,\cite{Klint2011}, relying on
the concrete semantics from Rascal Light\,\cite{DBLP:journals/corr/Al-Sibahi17}.

We consider algebraic data types ($\mathit{at}$) and finite sets ($\keyword{set}\ttuple{t}$) of elements of type $t$.  Each algebraic data type $\mathit{at}$ has a set of unique constructors.  Each constructor $k(\esequence{t})$ has a fixed set of typed parameters.  The language includes sub-typing, with $\keyword{void}$ and $\keyword{value}$ as bottom and top types respectively.
\[
  t \in \namedset{Type} \Coloneqq\; \keyword{void} \mid \keyword{set}\ttuple{t} \mid  \mathit{at} \mid \keyword{value}
\]

\noindent We consider the following subset of Rascal expressions: From left to right we have: variable access, assignments, sequencing, constructor expressions, set literal expressions, matching failure expression, and bottom-up visitors:
\begin{align*}
  e \Coloneqq & \; x \in \namedset{Var} \mid x = e \mid e ; e \mid k(\esequence{e}) \mid \{\esequence{e}\} \mid \keyword{fail} \mid \keyword{visit} \; e \;
  \esequence{\mathit{cs}}
  \\\mathit{cs} \Coloneqq & \; \keyword{case} \; p => e
\end{align*}
\noindent Visitors are a key construct in Rascal. A visitor $\keyword{visit} \; e \;
  \esequence{\mathit{cs}}$
traverses recursively the value obtained by evaluating $e$ (any combination of simple
values, data type values and collections). During the traversal,
case expression $\esequence{\mathit{cs}}$ are applied to the nodes,
and the values matching target patterns are rewritten. We will discuss a concrete subset of patterns $p$ further in \cref{sec:patternmatching}.
For brevity, we only discuss the bottom-up visitors in the paper. However, Rabit
(\cref{sec:evaluation}) supports all visitor strategies of Rascal.
\paragraph{Notation}
We write \hbox{$(x, y)\!\in\!f$} to denote the pair $(x, y)$ such that
\hbox{$x\!\in\!\text{dom}\,f$} and \hbox{$y\!=\!f(x)$}.
Abstract semantic components, sets, and operations are marked with a hat:
$\abst{a}$.  A sequence of $e_1,\dots,e_n$ is contracted using an underlining
$\esequence{e}$. The empty sequence is written by $\varepsilon$, and
concatenation of sequences $\esequence{e_1}$ and $\esequence{e_2}$ is written
$\esequence{e_1}, \esequence{e_2}$.  Notation is lifted to sequences in an
intuitive manner: for example  given a sequence $\esequence{v}$, the value $v_i$
denotes the $i$th element in the sequence, and $\esequence{v\!:\!t}$ denotes the
sequence $v_1\!:\!t_1, \dots, v_n\!:\!t_n$.


\section{Abstract Domains}
\label{sec:abstractdomains}

Our abstract domains are designed to allow modular composition.
Modularity is key for transformation languages, which manipulate a
large variety of kinds of values.
The design allows easily replacing abstract domains
for particular types of values, as well as adding support for new value types.
We want to construct an abstract value domain $\abst{\mathit{vs}} \in
\anamedset{ValueShape}$ which captures inductive refinement types of form:
\begin{gather*}
  \mathit{at}^r = k_1(\esequence{\abst{\mathit{vs}}_1}) \wr \dots
  \wr k_{\mathrm{n}}(\esequence{\abst{\mathit{vs}}_{\mathrm{n}}})
\end{gather*}
where each value $\abst{\mathit{vs}}_i$ can possibly recursively refer to
$\mathit{at}^r$. Below, we define abstract domains for sets, data
types and recursively defined domains.

The modular domain design generalizes parameterized
domains\,\cite{DBLP:conf/birthday/Cousot03} to follow a
design inspired by the modular construction of types and domains
\cite{DBLP:journals/siamcomp/Scott76%
  ,DBLP:conf/icfp/ChapmanDMM10,benkedybjerjansson2003:gendepty}.
The idea is to define domains parametrically---i.e. in the form \smash{$\anamedset{F}(\esequence{\anamedset{E}})$}---so that abstract domains for
subcomponents are taken as parameters, and explicit recursion is
handled separately. We use standard domain combinators\,\cite{DBLP:books/daglib/0070910/InformationSystems}
to combine the various domains into our target abstract value domain.

\paragraph{Set shape domain} Let $\namedset{Set}(\namedset{E})$ denote
the domain of sets consisting of elements taken from $\namedset{E}$. We define abstract finite sets using
abstract elements $\{\abst{e}\}_{[l;u]}$ from a parameterized domain
\smash{$\anamedset{SetShape}(\anamedset{E})$}.
The component from the parameter domain ($\abst{e} \in \anamedset{E}$) represents the abstraction of the shape of
elements, and a non-negative interval component $[l;u] \in
\anamedset{Interval\smash{^{+}}}$ is used to abstract over the cardinality
(so $l, u \in \mathbb{R}^{+}$ and $l \leq u$).
The abstract set element acts as a reduced product
between $\abst{e}$ and $[l;u]$ and thus the lattice operations follow directly.

 Given a concretization function for the abstract content domain
 $\gamma_{\anamedset{E}} \in \anamedset{E} -> \wp\left( \namedset{E}
 \right) $, we can define a concretization function for the abstract set shape domain
 to possible finite sets of concrete elements
 $\gamma_{\anamedset{SS}} \in \anamedset{SetShape}(\anamedset{E}) -> \wp\left(
   \namedset{Set}\left( \namedset{E} \right)  \right)$:
 \begin{gather*}
   \gamma_{\anamedset{SS}}(\{\abst{e}\}_{[l;u]}) = \left\{ \mathit{es} ~\middle|~
     \mathit{es} \subseteq \gamma_{\anamedset{E}}(\abst{e}) \wedge |\mathit{es}| \in \gamma_{\anamedset{I}}([l;u]) \right\}
 \end{gather*}

\begin{example}
  Let \anamedset{Interval}\ be a domain of intervals of integers (a standard abstraction over integers). We can concretize abstract elements from
  $\anamedset{SetShape}(\anamedset{Interval})$ to a set of possible sets of integers from $\wp\left( \namedset{Set}\left(
    \mathbb{Z}  \right) \right)$ as follows:
    $$\gamma_{\anamedset{SS}}(\{[42;43]\}_{[1;2]}) = \left\{ \{42\}, \{43\}, \{42,43\} \right\}$$
\end{example}

\paragraph{Data shape domain}
Inductive refinement types are defined as a generalization of refinement
types\,\cite{DBLP:conf/pldi/FreemanP91,DBLP:conf/pldi/XiP98,DBLP:journals/tse/RushbyOS98}
that inductively constrain the possible constructors and the content in a data structure.
We use a parameterized abstraction of data types
$\smash{\anamedset{DataShape}(\anamedset{E})}$, whose parameter
$\smash{\anamedset{E}}$ abstracts over the shape of constructor arguments:
\begin{multline*}
  \abst{d} {\in} \anamedset{DataShape}(\anamedset{E})  = \\
\{ \bot_{\anamedset{DS}} \} \cup
\{k_1(\esequence{e_1}) {\wr} \dots {\wr}
k_{\mathrm{n}}(\esequence{e_{\mathrm{n}}}) \mid e_i {\in} \anamedset{E}
\} \cup \{ \top_{\anamedset{DS}} \}
\end{multline*}%
We have the least element $\bot_{\anamedset{DS}}$ and top element
$\top_{\anamedset{DS}}$ elements---respectively representing no data
types value and all data type values---and otherwise a non-empty
choice between unique (all different) constructors of the same
algebraic data type $k_1(\esequence{e_1}) \wr \dots \wr
k_{\mathrm{n}}(\esequence{e_{\mathrm{n}}})$ (shortened
$\esequence{k}(\esequence{e})$).
We can treat the constructor choice as a finite map $[k_1 \mapsto \esequence{e_1},\dots,
k_{\mathrm{n}} \mapsto \esequence{e_{\mathrm{n}}}]$, and then directly define
our lattice operations point-wise.

Given a concretization function for the concrete content domain
$\gamma_{\anamedset{E}} \in \anamedset{E} -> \wp\left(
  \namedset{E} \right)$, we can create a concretization function for the data
shape domain $$\gamma_{\anamedset{DS}} \in \anamedset{DataShape}(\anamedset{E}) -> \wp\left(
  \namedset{Data}(\namedset{E}) \right)$$
where
$\namedset{Data}(\namedset{E}) = \left\{ k(\esequence{v}) ~\middle\vert~
    \exists \text{ a type }\mathit{at}.\ k(\esequence{v})\!\in\! \sem{\mathit{at}} \wedge
    \esequence{v \in \namedset{E}} \right\}$
The concretization is defined as follows:
\begin{gather*}
     \gamma_{\anamedset{DS}}(\bot_{\anamedset{DS}}) = \emptyset
  \qquad
  \gamma_{\anamedset{DS}}(\top_{\anamedset{DS}}) = \namedset{Data}(\namedset{E})  \\
  \gamma_{\anamedset{DS}}(k_1(\esequence{e_1}) \wr \dots \wr k_{\mathrm{n}}(\esequence{e_{\mathrm{n}}}))
  = \left\{ k_i(\esequence{v}) ~\middle|~ i \in [1, \mathrm{n}] \wedge
    \esequence{v \in \gamma_{\anamedset{E}}(e_i)} \right\}
\end{gather*}%

\begin{example}
  We can concretize abstract
  data elements $\anamedset{DataShape}(\anamedset{Interval})$ to a set of
  possible concrete data values $\wp\left( \namedset{Data}(\mathbb{Z}) \right)$.
  Consider values from the
  algebraic data type:
  \[\keyword{data} \; \auxiliary{errorloc} = \auxiliary{repl}() \mid
    \auxiliary{linecol}(\keyword{int}, \keyword{int}) \]
  We can concretize abstracting elements as follows:
  \begin{multline*}
    \hspace{10mm}\gamma_{\anamedset{DS}}(\auxiliary{repl}(){\,\wr\,}
    \auxiliary{linecol}([1;1], [3;4])) = \\
    \left\{\auxiliary{repl}(),
      \auxiliary{linecol}(1, 3), \auxiliary{linecol}(1, 4) \right\}\hspace{10mm}
  \end{multline*}
\end{example}%
\paragraph{Recursive shapes}

We extend our abstract domains to cover recursive structures such as
lists and trees.  Given a type expression $\namedset{F}(X)$ with a
variable $X$, we construct the abstract domain as the solution to the
recursive equation $X =
\namedset{F}({X})$\,\cite{DBLP:journals/siamcomp/Scott76,DBLP:journals/siamcomp/SmythP82,DBLP:books/daglib/0070910/InformationSystems}, obtained by iterating the induced map \namedset{F} over the empty
domain $\mathbb{0}$ and adjoining a new top element to the limit
domain. The concretization function of the recursive domain follows directly from the
concretization function of the underlying functor domain.

\begin{example}
  We can concretize abstract elements of the refinement type  from
  our running example:
  \begin{gather*}
    \gamma_{\anamedset{DS}}(\auxiliary{Expr}^{e}) =
    \left\{
      \begin{gathered}
{\overbrace{\auxiliary{cst}(\auxiliary{suc}(\auxiliary{suc}(\auxiliary{zero})))}^2},
\auxiliary{mult}(2, 2), \\ \auxiliary{mult}(\auxiliary{mult}(2, 2), 2), \dots
      \end{gathered}
\right\}
  \end{gather*}
  where $\auxiliary{Expr}^{e} =
  \auxiliary{cst}(\auxiliary{suc}(\auxiliary{suc}(\auxiliary{zero}))) \wr
  \auxiliary{mult}(\auxiliary{Expr}^{e}, \auxiliary{Expr}^{e})$
  In particular, our abstract element represents the set of all multiplications
  of the constant $2$.
\end{example}

\paragraph{Value domains}

We presented the required components for abstracting
individual types, and now all that is left is putting everything together.
We construct our value shape domain using choice and recursive
domain equations:
\begin{multline*}
\anamedset{ValueShape} = \\
\anamedset{SetShape}(\anamedset{ValueShape}) \oplus \anamedset{DataShape}(\anamedset{ValueShape})
\end{multline*}
Similarly, we have the corresponding concrete shape domain:
\[\namedset{Value} = \namedset{Set}\left( \namedset{Value} \right) \uplus \namedset{Data}(\namedset{Value}) \]
We then have a concretization function $\gamma_{\anamedset{VS}} \in
\anamedset{ValueShape} -> \wp\left( \namedset{Value} \right)$, which follows
directly from the previously defined concretization functions.

\subsection*{Abstract state domains}
We now explain how to construct abstractions of states and results when executing Rascal programs.

\paragraph{Abstract store domain}
Tracking assignments of variables is important since matching
 variable patterns depends on the value being assigned in the store:
\[
  \abst{\sigma} \in \anamedset{Store} = \namedset{Var} \rightarrow
  \{\auxiliary{ff}, \auxiliary{tt}\} \times  \anamedset{ValueShape}
\]

\noindent  For a variable $x$ we get $\abst\sigma (x) = (b,\abst{\mathit{vs}})$ where $b$ is true if $x$ might be unassigned, and false otherwise (when $x$ is definitely assigned). The second component, $\abst{\textit{vs}}$ is a shape approximating a possible value of $x$.

We lift the orderings and lattice operations point-wise from the value shape domain to
abstract stores. We define the concretization function $\gamma_{\anamedset{Store}} \in
\anamedset{Store} -> \wp\left(\namedset{Store}\right)$ as:
\begin{align*}
  \gamma_{\anamedset{Store}}(\abst{\sigma}) = \left\{\, \sigma ~\middle\vert~
  \begin{gathered}
    \forall x, b, \abst{\mathit{vs}}.\; \abst{\sigma}(x) = (b, \abst{\mathit{vs}}) => \\
    (\neg b => x\in\mathbf{dom}\;\sigma) \\ {}\wedge   (x \in \mathbf{dom} \; \sigma => \sigma(x) \in \gamma_{\anamedset{V}}(\abst{\mathit{vs}}))
  \end{gathered}
  \right\}
\end{align*}

\paragraph{Abstract result domain}
Traditionally, abstract control flow is handled using a collecting denotational
semantics with continuations, or by
explicitly constructing a control flow graph. These methods are
non-trivial to apply for a rich language like Rascal, especially considering backtracking,
exceptions and data-dependent control flow introduced by visitors.
A nice side-effect of Schmidt-style abstract interpretation
is that it allows handling abstraction of control flow directly.

We model different type of results---successes, pattern match failures, errors directly in a \smash{$\anamedset{ResSet}$} domain which keeps
track of possible results with each its own separate store. Keeping separate
stores is important to maintain
precision around different paths:
\begin{gather*}
  \mathit{rest} \in \anamedset{ResType} \Coloneqq \keyword{success} \mid \mathit{exres} \\
  \mathit{exres} \Coloneqq
  \keyword{fail} \mid \keyword{error} \quad
  \abst{\mathit{resv}} \in \anamedset{ResVal} \Coloneqq \cdot \mid \abst{\mathit{vs}} \\
  \abst{\mathit{Res}} \in \anamedset{ResSet} = \anamedset{ResType}
  \rightharpoonup \anamedset{ResVal} \times \anamedset{Store}
\end{gather*}
The lattice operations are lifted directly from the target value domains and
store domains.
We define the concretization function $\gamma_{\anamedset{RS}} \in
\anamedset{ResultSet} -> \wp\left( \namedset{Result} \times \namedset{Store}
\right)$:
\begin{gather*}
  \gamma_{\anamedset{RS}}(\abst{\mathit{Res}}) =
 \left\{
   (\mathit{rest} \; \mathit{resv}, \sigma) ~\middle\vert~
    \begin{gathered}
          \tuple{\mathit{rest},\tuple{\abst{\mathit{resv}}, \abst{\sigma}}} \in
    \abst{\mathit{Res}} \wedge{} \\ \mathit{resv} \in
    \gamma_{\anamedset{RV}}(\abst{\mathit{resv}}) \wedge \sigma \in \gamma_{\anamedset{Store}}(\abst{\sigma})
    \end{gathered}
    \right\}
  \end{gather*}

\vspace{-1mm}


\section{Abstract Semantics}
\label{sec:abstractsemantics}

A distinguishing feature of Schmidt-style abstract interpretation is
that the derivation of abstract operational rules from a given
concrete operational semantics is systematic and to a large extent
mechanisable\,\cite{DBLP:journals/lisp/Schmidt98,DBLP:conf/cpp/BodinJS15}.
The creative work is therefore reduced to providing abstract definitions for
conditions and semantic operations such as pattern matching, and defining
\emph{trace memoization strategies} for non-structurally recursive operational rules,
to finitely approximate an infinite number of concrete traces
and produce a terminating static analysis.

\begin{figure}[t]
  \centering
{
\setlength{\abovedisplayskip}{1.5em}
\setlength{\belowdisplayskip}{1.5em}
\vspace{1.5em}
\begin{gather*}
  \evalexpr{\tikz[remember picture,baseline=-0.5ex]{\node[minimum
      height=1.5em,fill=psred!50] (concrete-exp)
      {$e$};}}{\tikz[remember picture,baseline=-0.5ex]{\node[minimum
      height=1.5em,fill=psgreen!50]
      (concrete-store) {$\sigma$};}}{\tikz[remember picture,baseline=-0.5ex]{\node[minimum
      height=1.5em,fill=psblue!50]
      (concrete-res-val) {$\mathit{rest} \;
        \mathit{resv}$};}}{\tikz[remember picture,baseline=-0.5ex]{\node[minimum height=1.5em,fill=psblue!50]
      (concrete-res-store) {$\sigma'$};}} \qquad\qquad \aevalexpr{\tikz[remember picture,baseline=-0.5ex]{\node[minimum
      height=1.5em,fill=psred!50] (abstract-exp)
      {$e$};}}{\tikz[remember picture,baseline=-0.5ex]{\node[minimum
      height=1.5em,fill=psgreen!50]
      (abstract-store) {$\abst{\sigma}$};}}{\tikz[remember picture,baseline=-0.5ex]{\node[minimum
      height=1.5em,fill=psblue!50]
      (abstract-res) {$\abst{\mathit{Res}}$};}}
\end{gather*}
}
\begin{tikzpicture}[remember picture,overlay]
  \draw[->,psred] (abstract-exp) -- ++(0,1.5em) -| (concrete-exp) node[above,pos=.43] {\small same syntax};
  \draw[->,psgreen!120] (abstract-store) -- ++(0,-1.5em) -| (concrete-store) node[below,very near start] {\small abstracts input store};
    \draw[->,psblue] (abstract-res) -- ++(0,2em) -| (concrete-res-val) node[above,near start] {\small abstracts over sets of result values and stores};
  \draw[->,psblue] (abstract-res) -- ++(0,2em) -| (concrete-res-store);
\end{tikzpicture}
\vspace{-2mm}
\caption{Relating concrete semantics (left) to abstract semantics (right).}
\label{fig:relatingconcrabs}
\end{figure}
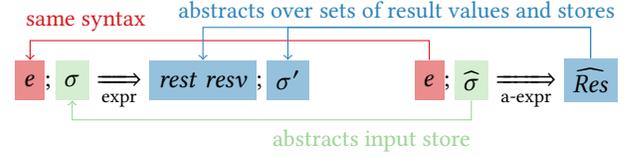

\Cref{fig:relatingconcrabs} relates the concrete evaluation
judgment (left) to the abstract
evaluation judgment (right) for Rascal expressions.
Both judgements evaluate the same expression $e$. The abstract evaluation
judgment abstracts the
initial concrete store $\sigma$ with an abstract store $\abst{\sigma}$.
The result of the abstract evaluation is a finite result set
$\abst{\mathit{Res}}$, abstracting over possibly infinitely many concrete result
values $\mathit{rest} \; \mathit{resv}$ and stores\nolinebreak\ $\sigma'\!$. $\abst{\mathit{Res}}$ maps each result type $\mathit{rest}$ to a pair of abstract result value
$\abst{\mathit{resv}}$ and abstract result store $\abst{\sigma}'$, i.e.:
\[
\abst{\mathit{Res}} =
[\mathit{rest}_1 \mapsto \tuple{\abst{\mathit{resv}_1},
    \abst{\sigma}_1},\dots,\mathit{rest}_{\mathrm{n}} \mapsto
  \tuple{\abst{\mathit{resv}_{\mathrm{n}}}, \abst{\sigma}_{\mathrm{n}}}]
\]

\noindent There is an important difference in how the concrete and
abstract semantic rules are used.
In a concrete operational semantics a language
construct is usually evaluated as soon as the premises of a
rule are satisfied. When evaluating abstractly, we must consider \emph{all}
applicable rules, to soundly over-approximate the possible concrete executions.
To this end, we introduce a special notation
to collect all derivations with the same input $i$ into a single derivation with output $O$ equal to the join of the individual outputs:
\[
  \collect{i \Rightarrow O} \triangleq O = \bigsqcup\left\{ o | i \Rightarrow o \right\}
\]

\noindent Let's use the operational rules for variable accesses to illustrate
the steps in Schmidt-style translation of operational rules.  The
concrete semantics contains two rules for variable accesses,
\textsc{E-V-S} for successful lookup, and \textsc{E-V-Er} for producing errors when accessing unassigned variables:
\begin{gather*}
  \inference[\textsc{E-V-S}]{x \in \textrm{dom }
    \sigma}{\evalexpr{x}{\sigma}{\keyword{success} \; \sigma(x)}{\sigma}} \\
  \inference[\textsc{E-V-Er}]{x \notin \textrm{dom }
    \sigma}{\evalexpr{x}{\sigma}{\keyword{error}}{\sigma}}
\end{gather*}

\noindent We follow three steps, to translate the concrete rules to abstract operational rules:

\begin{enumerate}
\item For each concrete rule, create an abstract rule that uses a judgment
  for evaluation of a syntactic form, e.g., $\textsc{AE-V-S}$ and
  $\textsc{AE-V-Er}$ for variables.
\item Replace the concrete conditions and semantic operations with the
  equivalent abstract conditions and semantic operations for target abstract
  values, e.g. $x \in \mathrm{dom} \; \sigma$ with $\abst{\sigma}(x) = \tuple{b,
    \abst{vs}}$ and a check on $b$. We obtain two execution rules:\\
\begin{gather*}
  \inference[\textsc{AE-V-S}]{{\abst{\sigma}(x) = (b, \abst{\mathit{vs}})}}{\aevalexprc{x}{\abst{\sigma}}{[\keyword{success} \mapsto
      \tuple{\abst{\mathit{vs}}, \abst{\sigma}}]}{v}}
 \\
 \inference[\textsc{AE-V-ER}]{\abst{\sigma}(x) =
   (\auxiliary{tt},
   \abst{\mathit{vs}})}{\aevalexprc{x}{\abst{\sigma}}{[\keyword{error} \mapsto \tuple{\cdot,\abst{\sigma}}]}{v}}
\end{gather*}
Observe when $b$ is true, both a success and failure may occur, and we need
rules to cover both cases.

\item Create a rule that collects all possible evaluations of the syntax-specific judgment rules, e.g. $\textsc{AE-V}$ for variables:
\begin{gather*}
  \inference[\textsc{AE-V}]
  {\collect{\aevalexprc{x}{\abst{\sigma}}{\abst{\mathit{Res}}'}{v}{}}}
  {\aevalexpr{x}{\abst{\sigma}}{\abst{\mathit{Res}}}'}
\end{gather*}%
\end{enumerate}

\noindent The possible shapes of the result value depend on the pair assigned to $x$ in the abstract
store. If the value shape of $x$ is $\bot$, we drop the success result from the
result set.
The following examples illustrate the possible outcome result shapes:%

\begin{center}
  \renewcommand{\arraystretch}{1.4}\small
  \begin{tabular}{lll}
    \textbf{Assigned Value~~~} & \textbf{Result Set} & \textbf{Rules} \\ \hline
    $\abst{\sigma}(x) = \tuple{\auxiliary{ff}, \bot_{\anamedset{VS}}}$ & $[]$ & \small\textsc{AE-V-S} \\
    $\abst{\sigma}(x) = \tuple{\auxiliary{ff}, [1;3]}$ & $[\keyword{success} \mapsto \tuple{[1;3], \abst{\sigma}}]$ & \small\textsc{AE-V-S} \\
    $\abst{\sigma}(x) = \tuple{\auxiliary{tt}, \bot_{\anamedset{VS}}}$ & $[\keyword{error} \mapsto \tuple{\cdot, \abst{\sigma}}]$ & \small\textsc{AE-V-S}, \textsc{AE-V-Er} \\[3mm]
    $\abst{\sigma}(x) = \tuple{\auxiliary{tt}, [1;3]}$ &  $\begin{gathered}
                                                           [\keyword{success} \mapsto \tuple{[1;3], \abst{\sigma}}, \\ \keyword{error} \mapsto \tuple{\cdot, \abst{\sigma}}]
                                                         \end{gathered}$~~ & \small\textsc{AE-V-S}, \textsc{AE-V-Er}   \\
  \end{tabular}
\end{center}

\smallskip

\noindent
It is possible to translate the
operational semantics rules for other basic
expressions using the presented steps (see \Cref{sec:abssemanticrules}).
The core changes are the ones moving from checks of definiteness to checks of \emph{possibility}.
For example:
\begin{itemize}

\item \sloppy Checking that evaluation of $e$ has
  succeeded, requires that the abstract semantics uses \smash{$\aevalexpr{e}{\abst{\sigma}}{\abst{\mathit{Res}}}$} and
  $\tuple{\keyword{success}, \tuple{\abst{\mathit{vs}}, \abst{\sigma}'}} \in  \abst{\mathit{Res}}$, as
  compared to \smash{$\evalexpr{e}{\sigma}{\keyword{success} \; v}{\sigma'}$} in the
  concrete semantics.

\item \strut Typing is now done using abstract judgments
  \smash{$\atyping{\abst{\mathit{vs}}}{t}$ and $\asubtyping{t}{t'}$}. In particular, type $t$ is an abstract
  subtype of type $t'$ ($\asubtyping{t}{t'}$) if there is a subtype $t''$ of $t$
  ($\subtyping{t''}{t}$) that is also a subtype of $t'$ ($\subtyping{t''}{t'}$).
  This implies that \smash{$\asubtyping{t}{t'}$ and $\anotsubtyping{t}{t'}$} are
  non-exclusive.

\item To check whether a particular constructor is possible, we use the
  abstract auxiliary function $\smash{\aauxiliary{unfold}}(\abst{\mathit{vs}}, t)$
  which produces a refined value of type $t$ if possible---splitting alternative
  constructors for data type values---and additionally produces $\keyword{error}$ if the value is
  possibly not an element of $t$.

\end{itemize}


\section{Pattern Matching} \label{sec:patternmatching}

Expressive pattern matching is key feature of high-level transformation
languages. Rabit handles the full Rascal pattern language including
type-based matching and deep pattern matching. For brevity, we discuss a
subset, including variables $x$, constructor patterns $k(\esequence{p})$, and set patterns $\{\esequence{{{\star}p}}\}$:
\begin{equation*}
  p \Coloneqq x \mid k(\esequence{p}) \mid \{\esequence{{{\star}p}}\} 
  \qquad
  {{\star}p} \Coloneqq p \mid {{\star}x}
\end{equation*}

\noindent Rascal allows non-linear matching where the same variable $x$ can be
mentioned more than once: all values matched against $x$ must have equal values
for the match to succeed.  Each set pattern contains a sequence of sub-patterns
${{\star}p}$; each sub-pattern in the sequence is either an ordinary pattern $p$ matched against a single set element, or a star pattern ${{\star}x}$ to be matched against a subset of elements.  Star patterns can backtrack when pattern matching fails because of non-linear variable references, or when explicitly triggered by the $\keyword{fail}$ expression.

This expressiveness poses challenges for developing an abstract
interpreter that is not only sound, but is also sufficiently
\emph{precise} to prove interesting properties.  The key
aspects of Rabit in handling pattern matching is how we
maintain precision by \emph{refining} input values on pattern matching
successes and failures.

\begin{figure*}[t]
\centering
\begin{subfigure}[t]{.43\linewidth}
\begin{mdframed}[align=center, userdefinedwidth=\textwidth]
{\small
  $\begin{aligned}
    k(\esequence{v}) & |=_\rho k(\esequence{p})
    & & \textbf{iff}
    & & \esequence{t} \text{ are parameter types of } k
    \\[-2mm]
    & & & & & \textbf{and }  \esequence{v : t'}
              \textbf{ and } \esequence{t' <: t} \\[-2mm]
    & & & & & \textbf{and }
    \satsemstar{\esequence{v}}{\rho}{\esequence{p}}
    \\
    v & |=_\rho x
    & & \textbf{iff}
    & & \rho(x) = v
    \\
    \{\esequence{v}\} & |=_\rho \{\esequence{{\star}p}\}
    & & \textbf{iff}
    & & \satsemstar{\esequence{v}}{\rho}{\esequence{{\star}p}}
    \\
    \varepsilon & |=_\rho^{\star} \varepsilon
    & & \textbf{always}
    \\
    v,\esequence{v'} & |=_\rho^{\star} p,\esequence{{\star}p'}
    & & \textbf{iff}
    & & \satsem{v}{\rho}{p} \textbf{ and }
    \satsemstar{\esequence{v'}}{\rho}{\esequence{{\star}p'}}
    \\
    \esequence{v},\esequence{v'} & |=_\rho^{\star} {\star}x,\esequence{{\star}p'}
    & & \textbf{iff}
    & & \rho(x) = \{\esequence{v}\} \textbf{ and } \satsemstar{\esequence{v'}}{\rho}{\esequence{{\star}p'}}
  \end{aligned}$}
  \vspace{-2mm}
\end{mdframed}
\caption{Concrete ($v |=_\rho p$ reads: $v$ matches $p$ with $\rho$)}
\label{fig:concrgensem}
\smallskip
  \end{subfigure}
  \quad
  \begin{subfigure}[t]{.53\linewidth}
\begin{mdframed}[align=center, userdefinedwidth=\textwidth]
{\small
  $\begin{aligned}
    k(\esequence{\abst{\mathit{vs}}}) & \;\abst{|=}_{\abst{\rho}}\; k(\esequence{p})
    & & \textbf{iff}
    & & \esequence{t} \text{ are parameter types of } k \\[-2mm]
    & & & & & \textbf{and }
      \esequence{\atyping{\abst{\mathit{vs}}}{t'}} \textbf{ and }
      \esequence{\asubtyping{t'}{t}} \textbf{ and }
      \abssatsemstar{\esequence{\abst{\mathit{vs}}}}{\abst{\rho}}{\esequence{p}}
    \\
    \abst{\mathit{vs}} & \;\abst{|=}_{\abst{\rho}}\; x
    & &  \textbf{iff}
    & & \abst{\rho}(x) \sqsubseteq \abst{\mathit{vs}}
    \\
    \{\esequence{\abst{\mathit{vs}}}\}_{[l;u]} & \;\abst{|=}_{\abst{\rho}}\; \{\esequence{{\star}p}\}
     & & \textbf{iff}
     & & \abssatsemstar{\abst{\mathit{vs}}, [l;u]}{\abst{\rho}}{\esequence{{\star}p}}
     \\
     \abst{\mathit{vs}}, [0;u] & \;\abst{|=}^{\star}_{\abst{\rho}}\; \varepsilon
     & & \textbf{always}
     \\
     \abst{\mathit{vs}}, [l;u] & \;\abst{|=}^{\star}_{\abst{\rho}} \; p, \esequence{{\star}p'}
     & & \textbf{iff}
     & & u > 0 \textbf{ and }
         \abssatsem{\abst{\mathit{vs}}}{\abst{\rho}}{p} \\[-2mm]
    & & & & & \textbf{and }  \abssatsemstar{\abst{\mathit{vs}}, [l-1;u-1]}{\abst{\rho}}{p,
       \esequence{{\star}p'}}
     \\
     \abst{\mathit{vs}}, [l;u] & \; \abst{|=}^{\star}_{\abst{\rho}} \; \star x,
     \esequence{{\star}p'}
     & & \textbf{iff}
    & & \abst{\rho}(x) = \{\abst{\mathit{vs}}'\}_{[l';u']} \\[-2mm]
    & & & & & \textbf{and } l' \leq l \textbf{ and } u' \leq u \textbf{ and } \abst{\mathit{vs}}' \sqsubseteq
   \abst{\mathit{vs}} \\[-2mm]
     & & & & & \textbf{and } \abssatsemstar{\abst{\mathit{vs}}, [l-u';u-l']}{\abst{\rho}}{\esequence{{\star}p'}}
  \end{aligned}$}
  \vspace{-2mm}
 \end{mdframed}
 \caption{Abstract ($\abst{vs}\;\abst{|=}_{\abst{\rho}}\;\abst{p}$ reads: $\abst{vs}$ may match $\abst p$ with $\abst \rho$)}
 \label{fig:absgensem}
  \end{subfigure}
  \vspace{-1mm}
  \caption{Satisfiability semantics for pattern matching}
  \label{fig:pattgensem}
  \bigskip
\end{figure*}

\subsection{Satisfiability semantics for patterns} We begin by defining what it means that a (concrete/abstract) value matches a pattern.  \Cref{fig:concrgensem} shows the concrete semantics for patterns. In the figure, $\rho$ is a binding environment:
\[
  \rho \in \namedset{BindingEnv} = \namedset{Var} \rightharpoonup \namedset{Value}
\]
A value $v$ matches a pattern $p$ ($\satsem{v}{}{p}$) iff there exists a binding environment $\rho$ that maps the variables in the pattern to values in $\textbf{dom} \, \rho = \auxiliary{vars}(p)$ so that $v$ is accepted by the satisfiability semantics $\satsem{v}{\rho}{p}$ as defined in \cref{fig:concrgensem}.

Constructor patterns $k(\esequence{p})$ accept any well-typed value $k(\esequence{v})$ of the same constructor whose subcomponents $\esequence{v}$ match the sub-patterns $\esequence{p}$ consistently in the same binding environment $\rho$. A variable $x$ matches exactly the value it is bound to in the binding environment $\rho$. A set pattern $\{\esequence{{\star}p}\}$ accepts any set of values $\{\esequence{v}\}$ such that an associative-commutative arrangement of the sub-values $\esequence{v}$ matches the sequence of sub-patterns $\esequence{{\star}p}$ under $\rho$.

A value sequence $\esequence{v}$ matches a pattern sequence
$\esequence{{\star}p}$ ($\satsemstar{\esequence{v}}{}{\esequence{{\star}p}}$) if there exists a binding environment $\rho$ such that
$\textbf{dom} \; \rho = \auxiliary{vars}(\esequence{{\star}p})$ and $\satsemstar{\esequence{v}}{\rho}{\esequence{{\star}p}}$. An empty
sequence of patterns $\varepsilon$ accepts an empty sequence of values
$\varepsilon$. A sequence starting $p,\esequence{{\star}p'}$ with an ordinary
pattern $p$ matches any non-empty sequence of values $v,\esequence{v'}$ where
$v$ matches $p$ and $\esequence{v'}$ matches
$\esequence{{\star}p'}$ consistently under the same binding environment $\rho$.
A sequence ${\star}x,\esequence{{\star}p'}$ works analogously but it splits the
value sequence in two $\esequence{v}$ and $\esequence{v'}$, such that $x$ is
assigned to $\esequence{v}$ in $\rho$ and $\esequence{v'}$ matches
$\esequence{{\star}p'}$ consistently in $\rho$.

\begin{example}
  We revisit the running example to understand how the data type values are matched. We consider matching the following set of expression values:
 \[
    \{\overbrace{\mult{\cst{\zero}, \cst{\suc{\zero}}}, \cst{\zero}}^{\esequence{v}}\}
  \]
  against the pattern {$p=\{\mult{x,y}, {{\star}w}, x\}$} in the environment {$\rho = [x \mapsto \cst{\zero},y \mapsto \cst{\suc{\zero}}, w\mapsto\{\}]$}. The matching argument is as follows:
  \begin{flalign*}
      & \hspace{\parindent}\satsem{\{\esequence{v}\}}{\rho}{p}\textbf{ iff } \satsemstar{\esequence{v}}{\rho}{\mult{x,y},{\star}w,x}
      \\[-1mm]
      & \hspace{\parindent}\textbf{iff }\satsem{\mult{\cst{\zero},
          \cst{\suc{\zero}}}}{\rho}{\mult{x,y}} \\
      & \hspace{\parindent}\textbf{and}~ \satsemstar{\cst{\zero}}{\rho}{{\star}w,x} &
    \end{flalign*}
  We see that the first conjunct matches as follows:
  \begin{flalign*}
      & \hspace{\parindent}\satsem{\mult{\cst{\zero}, \cst{\suc{\zero}}}}{\rho}{\mult{x,y}}
      \\[-1mm]
      & \hspace{\parindent}\textbf{iff }\satsemstar{\cst{\zero}, \cst{\suc{\zero}}}{\rho}{x,y}
      \\[-1mm]
      & \hspace{\parindent}\textbf{iff }\rho(x) = \cst{\zero} ~\textbf{and}~ \rho(y) = \cst{\suc{\zero}} &
    \end{flalign*}
Similarly, the second matches as follows:
  \begin{flalign*}
      & \hspace{\parindent}\satsemstar{\cst{\zero}}{\rho}{{\star}w,x}  \textbf{ iff }\rho(w) = \{\} ~\textbf{and}~ \rho(x) = \cst{\zero} &
    \end{flalign*}
\end{example}

\noindent The abstract pattern matching semantics (\cref{sub@fig:absgensem}) is analogous, but with a few noticeable differences. First, an abstract value $\abst{\mathit{vs}}$ matches a pattern $p$ ($\abssatsem{\abst{\mathit{vs}}}{}{p}$) if there exists a more precise value $\abst{\mathit{vs}}'$ (so $\abst{\mathit{vs}}' \sqsubseteq \abst{\mathit{vs}}$) and an abstract binding environment $\abst{\rho}$ with $\textbf{dom}\;\abst{\rho} = \auxiliary{vars}(p)$ so that $\abssatsem{\abst{\mathit{vs}}'}{\abst{\rho}}{p}$. The reason for using a more precise shape is the potential loss of information during over-approximation---a more precise value might have matched the pattern, even if the relaxed value does not necessarily.  Second, sequences are abstracted by shape--lengths pairs, which needs to be taken into account by sequence matching rules.  This is most visible in the very last rule, with a star pattern ${\star}x$, where we accept any assignment to a set abstraction $\abst{\mathit{vs}}$ which has a more precise shape and a smaller length.

\subsection{Computing pattern matches}

The declarative satisfiability semantics of patterns, albeit quite clean, is
unfortunately not directly computable. In Rabit, we rely on an abstract
operational semantics (see \cref{sec:oppatts}), translated from the concrete operational pattern matching
semantics\,\cite{DBLP:journals/corr/Al-Sibahi17}, using similar technique to the
one presented in \cref{sec:abstractsemantics}. The interesting ideas are in the
refining semantic operators used, which we will discuss further.

\paragraph{Semantic operators with refinement}
Since Rascal supports non-linear matching, it becomes necessary to merge
environments computed when matching sub-patterns to check whether a match
succeeds or not. In abstract interpretation, we can refine the abstract
environments when merging for each possibility.
Consider when merging two abstract environments, where some variable $x$ is
assigned to $\abst{\mathit{vs}}$ in one, and $\abst{\mathit{vs}}'$ in the other.
If $\abst{\mathit{vs}}'$ is possibly equal to
$\abst{\mathit{vs}}$, we refine both values using this equality
assumption $\abst{\mathit{vs}} ~\abst{=}~ \abst{\mathit{vs}}'$.
Here, we have that abstract equality is defined as the
greatest lower bound if the value is non-bottom, i.e. $\abst{\mathit{vs}}
~\abst{=}~ \abst{\mathit{vs}}' \triangleq \{  \abst{\mathit{vs}}'' | \abst{\mathit{vs}}'' = \abst{\mathit{vs}}
\sqcap \abst{\mathit{vs}}' \neq \bot \}$.
Similarly, we can also refine both values if they are possibly non-equal
$\abst{\mathit{vs}} ~\abst{\neq}~ \abst{\mathit{vs}}'$.
Here, abstract inequality is defined using relative complements:
\[\abst{\mathit{vs}}
  ~\abst{\neq}~ \abst{\mathit{vs}}' \triangleq
  \begin{gathered}
  \left\{  \tuple{\abst{\mathit{vs}}'', \abst{\mathit{vs}}'} |
\abst{\mathit{vs}}'' = \abst{\mathit{vs}} \setminus (\abst{\mathit{vs}}
\sqcap \abst{\mathit{vs}}') \neq \bot \right\} \cup{} \\ \left\{  \tuple{\abst{\mathit{vs}}, \abst{\mathit{vs}}''} |
\abst{\mathit{vs}}'' = \abst{\mathit{vs}}' \setminus (\abst{\mathit{vs}}
\sqcap \abst{\mathit{vs}}') \neq \bot \right\}
  \end{gathered}
\]

\noindent In our abstract domains, the relative complement ($\setminus$) is limited. We heuristically define it for interesting cases, and otherwise it degrades to identity in the first argument (no refinement).  There are however useful cases, e.g., for excluding unary constructors $\suc{\Nat} \wr \zero \setminus \zero = \suc{\Nat}$ or at the end points of a lattice $[1;10] \setminus [1;2] = [3;10]$.

Similarly, for matching against a constructor pattern $k(\esequence{p})$, the core idea is that we should be able to
partition our value space into two: the abstract values that match the
constructor and those that do not.
For those values that possibly match $k(\esequence{p})$, we produce a refined value with $k$ as
the only choice, making sure that the sub-values in the result are refined by
the sub-patterns $\esequence{p}$.

Otherwise, we exclude $k$ from the refined value.
For a data type abstraction exclusion
removes the pattern constructor from the possible choices
\begin{equation*}\aauxiliary{exclude}(k(\abst{\mathit{vs}}) \wr k_1(\abst{\mathit{vs}}_1) \wr
\dots \wr k_{\mathrm{n}}(\abst{\mathit{vs}}_{\mathrm{n}}) , k) = k_1(\abst{\mathit{vs}}_1) \wr
\dots \wr k_{\mathrm{n}}(\abst{\mathit{vs}}_{\mathrm{n}})
\end{equation*}
and does not change the input shape otherwise.


\section{Traversals}
\label{sec:traversals}

First-class traversals are a key feature of high-level transformation
languages, since they enable effectively transforming large abstract
syntax trees.
We will focus on the challenges for bottom-up traversals, but
they are shared amongst all strategies supported
in Rascal.
The core idea of a bottom-up traversal of an abstract value $\abst{\mathit{vs}}$, is to first
traverse children of the value
$\aauxiliary{children}(\abst{\mathit{vs}})$ possibly rewriting them, then
reconstruct a new value using the rewritten
children and finally traversing the reconstructed value.
The main challenge is handling traversal of children, whose
representation and thus execution rules depend on the particular abstract value.

Concretely, the $\aauxiliary{children}(\abst{\mathit{vs}})$ function returns a
set of pairs $(\abst{\mathit{vs}}', \abst{\mathit{cvs}})$ where the first component $\abst{\mathit{vs}}'$ is a refinement of $\abst{\mathit{vs}}$ that matches the shape of children $\abst{\mathit{cvs}}$ in the second component. For data type values the representation of children is a heterogeneous sequence of abstract values
$\esequence{\abst{\mathit{vs}}''}$, while for set values (and the top element)
the representation of
children is a pair $\tuple{\abst{\mathit{vs}}'', [l;u]}$ with the first
component representing the shape of elements and the second representing their
count. For example,
\begin{align*}
&\aauxiliary{children}(\mult{\Expr, \Expr} \wr
  \cst{\suc{\Nat}}) = \\ &\qquad \left\{
    \begin{gathered}
    \tuple{\mult{\Expr, \Expr}, (\Expr, \Expr)}, \\
    \tuple{\cst{\suc{\Nat}}, \suc{\Nat}}
    \end{gathered}
  \right\}
\end{align*}
and $\aauxiliary{children}(\{\Expr\}_{[1;10]}) = \{\tuple{\{\Expr\}_{[1;10]},
    \tuple{\Expr,{[1;10]}}}\}$.
Note how the $\aauxiliary{children}$ function maintains precision by
partitioning the
alternatives for data-types, when traversing each corresponding
sequence of value shapes for the children.

\paragraph{Traversing children}
The shape of execution rules depend on the
representation of children; this is consistent with the
requirements imposed by Schmidt\,\cite{DBLP:journals/lisp/Schmidt98}.
For heterogeneous sequences of value shapes $\esequence{\abst{\mathit{vs}}}$, the execution rules iterate through the sequence
recursively traversing each element.
Due to over-approximation we may re-traverse the same or a more precise value on
recursion, and so we need to use trace memoization (\cref{sec:memoizationstrategies}) to terminate.
For example the children of an expression $\Expr$ refer to itself:
\[
  \aauxiliary{children}(\Expr) = \left\{  \begin{gathered}
    \tuple{\mult{\Expr, \Expr}, (\Expr, \Expr)}, \\
    \tuple{\cst{\Nat}, {\Nat}},
    \tuple{\var{\str}, {\str}}
    \end{gathered} \right\}
\]
Traversing children represented by a shape-length pair, is directed by the
length interval $[l;u]$. If $0$ is a possible value of the length interval, then
traversal can finish, refining the input shape to be empty.
Otherwise, we perform another traversal recursively on the shape of elements and
recursively on a new shape-length pair which decreases the length, finally
combining their values. Note, that if the length is
unbounded, e.g. $[0;\infty]$, then the value can be decreased forever and trace
memoization is also needed here for termination. This means that trace
memoization must here be nested breadth-wise (when recursing on an unbounded
sequence of children), in addition to depth-wise (when recursing on children);
this can be computationally expensive, and we will discuss in
\cref{sec:evaluation} how our implementation handles that.


\section{Trace Memoization}
\label{sec:memoizationstrategies}
Abstract interpretation and static program analysis in general perform
fixed-point calculation for analysing unbounded loops
and recursion. In Schmidt-style abstract
interpretation, the main technique
to handle recursion is \emph{trace memoization}\,\cite{DBLP:journals/lisp/Schmidt98,DBLP:journals/corr/Rosendahl13}.
The core idea of \emph{trace memoization} is to detect
non-structural re-evaluation of the same program element, i.e.,
when the evaluation of a program element is
recursively dependent on itself, like a
$\keyword{while}$-loop or traversal.

The main challenge when recursing over inputs from infinite domains, is to
determine \emph{when} to merge recursive paths together to correctly
over-approximate concrete executions.
We present an extension that is still terminating, sound
and, additionally, allows calculating results with good precision.
The core idea is to partition the infinite input domain using a finite domain of
elements, and on recursion degrade input values using previously met input
values from the same partition. We assume that all our domains are
lattices with a widening operator. Consider a
recursive operational semantics judgment $i \Longrightarrow o$, with
$i$ being an input from domain $\anamedset{Input}$, and $o$ being the
output from domain $\anamedset{Output}$.  For this judgment, we
associate a memoization map $\abst{M} \in \anamedset{PInput} ->
\anamedset{Input} \times \anamedset{Output}$ where
$\anamedset{PInput}$ is a finite partitioning domain that has a Galois
connection with our actual input, i.e. $\anamedset{Input}
\galois{\alpha_{\abst{PI}}}{\gamma_{\abst{PI}}} \anamedset{PInput}$.
The memoization map keeps track of the previously seen input and
corresponding output for values in the partition domain.  For example,
for input from our value domain $\anamedset{Value}$ we can use the
corresponding type from the domain $\namedset{Type}$ as input to the
memoization map.\footnote{Provided that we bound the
  depth of type parameters of collections.} So for values $1$ and
$[2;3]$ we would use $\keyword{int}$, while for
$\auxiliary{mult}(\namedset{Expr}, \namedset{Expr})$ we would
use the defining data type $\namedset{Expr}$.

\noindent
We perform a fixed-point calculation over the evaluation of input $i$.
Initially, the memoization map $\abst{M}$ is $\lambda \mathit{pi}. \tuple{\bot, \bot}$, and during evaluation
  we check whether there was already a value from the same
  partition as $i$, i.e., $\alpha_{\abst{PI}}(i) \in \mathrm{dom} \; \abst{M}$.
  At each iteration, there are then two possibilities:%
  \begin{description}
  \item[Hit] The corresponding input partition key is in the
    memoization map and a less precise input is stored, so
    $\abst{M}(\alpha_{\abst{PI}}(i)) = \tuple{i', o'}$ where $i
    \sqsubseteq_{\anamedset{Input}} i'$. Here, the output
    value $o$ that is stored in the memoization map is returned as result.
  \item[Widen] The corresponding input partition key is in the memoization map, but an
    unrelated or more precise input is stored, i.e.,
    $\abst{M}(\alpha_{\abst{PI}}(i)) = \tuple{i'', o''}$ where $i
    \not\sqsubseteq_{\anamedset{Input}} i''$. In this case we continue evaluation
    but with a widened input $i' = i'' \nabla_{\anamedset{Input}} (i'' \sqcup i)$ and an
    updated map $\abst{M}' = [\alpha_{\abst{PI}}(i) \mapsto \tuple{i',
      o_{\mathrm{prev}}}]$. Here, $ o_{\mathrm{prev}}$ is the output of the
    last iteration for the fixed-point calculation for input $i'$, and is
    assigned $\bot$ on the initial iteration.
  \end{description}

\noindent
Intuitively, the technique is terminating because the partitioning is finite,
and widening ensures that we reach an upper bound of possible inputs in a finite
number of steps, eventually getting a hit. The fixed-point iteration also uses
widening to calculate an upper bound, which similarly finishes in a number of steps.
The technique is sound because we only use output for previous input that is
less precise; therefore our function is continuous and a
fixed-point exists.


\section{Experimental Evaluation} \label{sec:evaluation}

We demonstrate the ability of Rabit to verify type and inductive shape properties, using
five transformation programs across various applications. Three programs are classic examples, and two are
extracted from open source projects.

\paragraph{Negation Normal Form}\,(\textbf{NNF}) transformation\,\cite[Section
2.5]{DBLP:books/daglib/0022394} is a classical rewrite of a propositional
formula to combination of conjunctions and disjunctions of literals, so
negations appear only next to atoms. An implementation of this transformation should guarantee the following:
\begin{description}
\item[P1]  Implication is not used as a connective in the result
\item[P2] All negations in the result are in front of atoms
\end{description}
\paragraph{Rename Struct Field} (\textbf{RSF}) refactoring changes the name of a
field in a struct, and that all corresponding field access expressions are renamed correctly as well:
\begin{description}
\item[P3] Structure should not define a field with the old field name
\item[P4] No field access expression to the old field
\end{description}

\paragraph{Desugar Oberon-0} (\textbf{DSO}) transformation\,\cite{wirth1996compiler,DBLP:journals/scp/BastenB0KLLPSV15}, translates
for-loops and switch-statements to while-loops and nested if-statements,
respectively.
\begin{description}
\item[P5] \textsf{for} should be correctly desugared to \textsf{while}
\item[P6] \textsf{switch} should be correctly desugared to \textsf{if}
\item[P7] No auxiliary data in output
\end{description}

\paragraph{Code Generation for Glagol} (\textbf{G2P}) a DSL for REST-like web development, translated to PHP for execution.\footnote{\url{https://github.com/BulgariaPHP/glagol-dsl}} We are interested in the part of the generator that translates Glagol expressions to PHP, and the following properties:
\begin{description}
\item[P8] Output only simple PHP expressions for simple Glagol expression inputs
\item[P9] No unary PHP expressions if no sign marks or negations in Glagol input
\end{description}

\paragraph{Mini Calculational Language} (\textbf{MCL}) a programming
language text-book\,\cite{sestoft_hallenberg_2017} implementation of a small
expression language, with arithmetic and logical expressions, variables,
$\keyword{if}$-expressions, and $\keyword{let}$-bindings. The implementation
contains an expression simplifier (larger version of running example in
\cref{fig:runningexample}), a type inference procedure, an interpreter and a compiler.
\begin{description}
\item[P10] Simplification procedure produces a simplified expression with no
  additions with 0, multiplications with 1 or 0, subtractions with 0, logical
  expressions with constant Boolean operands, and \keyword{if}-expressions with
  constant Boolean conditions.
\item[P11] Arithmetic expressions with no variables have type $\keyword{int}$
  and no type errors
\item[P12] Interpreting expressions with no integer constants and
  $\keyword{let}$'s gives only Boolean values
\item[P13] Compiling expressions with no $\keyword{if}$'s produces no
  $\keyword{goto}$'sand $\keyword{if}$ instructions
\item[P14] Compiling expressions with no $\keyword{if}$'s produces no labels and does not change label counter
\end{description}

\noindent
All these transformations satisfy the following criteria:
\begin{enumerate}
\item They are formulated by an independent source,
\item They can be translated in relatively straightforward manner to our subset
  of Rascal, and
\item They exercise important constructs, including visitors and the expressive pattern matching
\end{enumerate}
We have ported all these programs to Rascal Light.

\paragraph{Threats to validity.} The programs are not selected randomly, thus it
is hard to generalize the results for other transformations. We mitigated this
by selecting transformations that are realistic and vary in authors, programming
style and purpose. While translating the programs to Rascal Light, we strived to
minimize the amount of changes, but generally bias cannot be ruled out entirely.

\paragraph{Implementation.}
We have implemented the abstract interpreter in a prototype tool, Rabit, for all of
Rascal Light following the
process described in
\cref{sec:abstractsemantics,sec:patternmatching,sec:traversals,sec:memoizationstrategies}.
This required handling additional aspects, not discussed in the paper:
\begin{enumerate}
\item Possibly undefined values
\item Extended result state with
more Control flow constructs, backtracking, exceptions, loop
control, and
\item Fine-tuning memoization strategies to the different looping constructs and recursive calls
\end{enumerate}
By default, we use
the top element $\top$ for the types specified as input.
The user can specify the initial data-type refinements, store and parameters, to
get a more precise result for target function to be abstractly interpreted.
The output of the tool is the abstract result value set of abstractly
interpreting target function, the resulting store state
and the set of relevant inferred data-type refinements.



The implementation extends standard regular tree grammar
operations\,\cite{DBLP:conf/fpca/AikenM91,DBLP:conf/fpca/CousotC95}, to handle
the recursive equations for the expressive abstract domains, including base
values, collections and heterogeneous data types.  We use a more precise
partitioning strategy for trace memoization when needed, which also takes the
set of available constructors into account for data types.  The source code of
our implementation, including subject transformations, is freely available.
\footnote{
  \url{https://github.com/itu-square/Rascal-Light}
}

\begin{table}[bp] 
  \centering\small
  \begin{tabular}{llrrcc}
    & Transformation~      & ~LOC   & ~Runtime\,[s]          & ~~~Property~ & ~Verified~ \\ \toprule
    \rowcolor{gray!10}
    &                   & &                     & P1       & \cmark   \\
    \rowcolor{gray!10}
    & \multirow{-2}{*}{NNF} & \multirow{-2}{*}{15} & \multirow{-2}{*}{7.3} & P2       & \cmark   \\
    &                   &    &                      & P3       & \xmark   \\
    & \multirow{-2}{*}{RSF}  & \multirow{-2}{*}{35} & \multirow{-2}{*}{6.0}  & P4       & \cmark   \\ \rowcolor{gray!10}
     &                      & &                      & P5       & \cmark   \\ \rowcolor{gray!10}
      &                    & &                      & P6       & \cmark   \\ \rowcolor{gray!10}
    & \multirow{-3}{*}{DSO} & \multirow{-3}{*}{125} & \multirow{-3}{*}{25.0} & P7       & \xmark   \\
    &                       & & 1.6                    & P8       & \cmark   \\
    & \multirow{-2}{*}{G2P} & \multirow{-2}{*}{350} & 3.5                   & P9       & \cmark   \\
    \rowcolor{gray!10}
    & & & 1.6 & P10 & \cmark \\
    \rowcolor{gray!10}
    & & & 0.7 & P11 & \cmark \\
    \rowcolor{gray!10}
    & & & 0.6 & P12 & \cmark \\
    \rowcolor{gray!10}
    & & &     & P13 & \cmark \\
    \rowcolor{gray!10}
    & \multirow{-5}{*}{MCL} & \multirow{-4}{*}{298} & \multirow{-2}{*}{0.9} & P14 & \cmark  \\
    \bottomrule
  \end{tabular}
  \caption{Time and success rate for analyzing programs and properties presented
    earlier this section. Time is the median of five runs.
    If the same time is reported for multiple properties, then they could be
    verified on the same input}
  \label{tab:rascalabsintresults}
  \vspace{-2mm}
\end{table}

\paragraph{Results.} We ran the experiments using Scala 2.12.2 on a 2012 Core i5
MacBook Pro.  \Cref{tab:rascalabsintresults} summarizes the size of the
programs, the runtime of the abstract interpreter, and whether the properties
have been verified.
Since we verify the results on the abstract shapes, the programs then are shown
to be correct for all possible concrete inputs satisfying the given properties.
We remark that all programs use the high-level expressive features of Rascal and
are thus significantly more succinct than comparable code in general purpose
languages.

The runtime, varying from single seconds to less than a minute, is reasonable.
All, but two, properties were successfully verified. The reason that our tool
runs slower on the DSO transformation than those with more lines of code (G2P
and MCL), is that it contains many nested traversals expressed as function
calls: our analysis is interprocedural but handles function calls by inlining
which can lead to some overhead during analysis.

Lines 1--2 in \cref{fig:nnfshape} show the input refinement type \namedset{FIn} for the normalization
procedure. The inferred inductive output type \namedset{FOut}
(lines 4--5) specifies that the implication is not present in the output (P1),
and negation only allows atoms as subformulae (P2). In fact, Rabit inferred a precise characterization of negation normal form as an inductive data type.

\begin{figure}[t]

  \caption{Initial and inferred refinement types for NNF}
\begin{lstlisting}[language=Rascal,xleftmargin=0cm]
  data FIn = and(FIn, FIn) | atom(str) | neg(FIn)
               | imp(FIn, FIn) | or(FIn, FIn)

  data FOut = and(FOut, FOut) | atom(str)
               | neg(atom(str)) | or(FOut, FOut)
\end{lstlisting}
  \label{fig:nnfshape}
\end{figure}

\section{Related Work}\label{sec:relatedwork}
We start with discussing techniques that could be used to make Rabit infer more
precise shapes and verify properties like P3 and P7.
To verify P3, we need to be able to relate field names to their corresponding
definitions in the field definition map of a class, which is not possible using
the presented non-relational abstract domains.
Relational abstract interpreteration\,\cite{DBLP:conf/pdo/MycroftJ85} allows
specifying such constraints that relate values across different variables, and even
inside and across substructures\,\cite{DBLP:conf/popl/ChangR08,DBLP:conf/pldi/HalbwachsP08,DBLP:journals/cl/LiuR17}.
For a concrete input of P7, we know that the number of auxiliary data elements
decreases on each iteration, but this information is lost in our
abstraction of data structures.
A possible solution could be to allow \emph{abstract attributes} that extract additional information
about the abstracted structures\,\cite{DBLP:conf/popl/SuterDK10,DBLP:conf/vmcai/BouajjaniDES12,DBLP:conf/vstte/PhamW13}.
For P7, a generalization of the multiset
abstraction\,\cite{DBLP:conf/vmcai/PerrelleH10} for data types, could be useful
to track e.g., the auxiliary statement count, and show that they decrease using
multiset-ordering\,\cite{DBLP:journals/cacm/DershowitzM79} like in term rewriting.
Other
techniques\,\cite{DBLP:conf/popl/ChangR08,DBLP:conf/esop/VazouRJ13,DBLP:conf/esop/AlbarghouthiBCK15}
support inferring inductive relational properties for general data-types---e.g,
binary tree property---but require a pre-specified structure to indicate the
places where refinement can happen.

Cousot and Cousot\,\cite{DBLP:conf/cc/CousotC02} present a general framework for modularly
constructing program analyses, but it requires a language with a compositional
control flow which Rascal does
not have.
Toubhans, Rival and
Chang\,\cite{DBLP:conf/vmcai/ToubhansCR13,DBLP:conf/isola/RivalTC14} develop a modular domain
design for pointer-manipulating programs supporting a rich set of fixed data
abstractions, whereas our domain construction focuses on providing automated
inference of inductive refinement types based on pure heterogeneous data-structures.

There are similarities between our work and verification techniques based on
program transformation (e.g.,
\cite{DBLP:journals/scp/AngelisFPP14,DBLP:journals/corr/LisitsaN15}) like
\emph{partial evaluation}\,\cite{DBLP:books/daglib/0072559} and
\emph{supercompilation}\,\cite{DBLP:journals/jfp/SorensenGJ96}. Our systematic
exploration of execution rules for abstraction is similar to \emph{unfolding},
and our use of widening is similar to \emph{folding}. The main difference
between the two techniques is that abstract interpretation mainly focuses on
capturing rich domains and performing widening at syntactic program points,
whereas program transformation based techniques often rely on symbolic inputs
and perform folding dynamically on the semantic execution graph during specialization.
We believe that there could benefits for the communities, to explore
combinations of these two approaches in the future.

Definitional interpreters have been suggested as a technique for building
compositional abstract interpreters\,\cite{DBLP:journals/pacmpl/DaraisLNH17}.
The idea is to rely on a monad transformer stack to share the implementation
of the concrete and abstract interpreters. We believe that our interpreter would
benefit by being written in such style\footnote{We only learned about
  this related work at a late stage}, which complements
our modular domain construction well. To ensure termination they rely on a
\emph{caching} algorithm, similar to ordinary finite input \emph{trace
  memoization}\,\cite{DBLP:journals/corr/Rosendahl13}.
Similarly, Van Horn and Might\,\cite{DBLP:conf/icfp/HornM10} present a systematic framework
to abstract higher-order functional languages with effects and complex control flow.
They rely on store-allocated continuations within abstract machines to handle recursion, which is then
kept finite during abstraction to ensure a terminating analysis.
Our technique focused on providing a more precise widening based on the
abstract input value, which was necessary for verifying the required properties
in our evaluation. We believe that it could be useful to look into abstract
machine-based abstractions in the future, in the case that higher-order
transformation languages need to be handled.

Garrigue\,\cite{garrigue1998programming,garrigue2004typing} presents algorithms for typing pattern
matching on polymorphic variant types in OCaml, where the set of constructors for a data type is not
fixed in advance. The theory is useful since it supports inferring simple
recursive shapes of programs, but it has its limitations: inference is
syntactic and exact, and it is
unclear how to generalize it to work with the rich
pattern matching constructs and heterogeneous visitors.
Haskell supports analysing coverage of its pattern matching language, that includes generalized algebraic
data types (GADTs) and Boolean constraints\,\cite{DBLP:conf/icfp/KarachaliasSVJ15}.
While general Haskell function calls can occur in the Boolean constraints, the
analysis treats them shallowly as function symbols; some covering pattern matches that depend on particular semantics of called
functions, will be marked falsely as non-exhaustive.
Modern SMT solvers supports reasoning with inductive functions defined over
algebraic data-types\,\cite{DBLP:conf/vmcai/ReynoldsK15}. The properties they
can verify are very expressive, and include inductive semantic properties.
The exact techniques employed are not very scalable, and encoding a
complex transformation directly would not finish verifying even simple
properties within reasonable time.  
Possible constructor analysis\,\cite{DBLP:conf/icfem/AndreescuJL15} has been used
to calculate the actual dependencies of a predicate and make flow-sensitive
analyses more precise. This is a type of shape
analysis that works with complex data-types and arrays, but only captures the prefix of the target structures.

Techniques for model transformation
verification based on static analysis\,\cite{DBLP:journals/tse/CuadradoGL17}
have been suggested, but are currently
focused on verification of rule errors based on types and undefinedness.
Symbolic execution has previously been
suggested\,\cite{DBLP:conf/sle/Al-SibahiDW16} as a way to validate high-level
transformation programs. However, that work
targets test generation rather than verification of properties.
Semantic typing\,\cite{DBLP:conf/icfp/CastagnaN08,DBLP:conf/popl/BenzakenCNS13} has been used to infer recursive type and shape properties for language with high-level constructs for querying and iteration.
The languages considered are however small calculi compared
to the supported subset of Rascal we consider, and our evaluation is significantly more extensive.


\section{Conclusion} \label{sec:conclusion}

Our goal was to use abstract interpretation to give a solid semantic foundation
for analyzing programs in modern high-level transformation languages.  To this
end we have designed and formalized a Schmidt-style abstract interpreter,
including \emph{partition-driven trace memoization} which works with infinite
input domains.
This worked well for a language like Rascal with complex control flow, and can
be adapted work for similar languages that have an operational semantics.
The proposed modular construction of abstract domains was vital for handling a language of this scale and complexity.

We implemented the interpreter as a tool, Rabit, which supports a non-trivial subset of Rascal, containing key features: several traversal
strategies, expressive pattern matching, backtracking, exceptions and control
operators, and generalized looping constructs.  We evaluated Rabit on classical transformations and on examples selected from open source projects, showing it allows verification of a series of sophisticated type and shape properties for these transformations.


\appendix
\section{Operational Pattern Matching}
\label{sec:oppatts}
\noindent

\begin{figure*}[!h]
  \centering
\vspace{2em}
  \begin{minipage}{1.0\linewidth}
\begin{gather*}
  \match{\tikz[remember picture,baseline=-0.5ex]{\node[minimum
      height=2em,fill=psred!50] (concrete-patt)
      {$p$}}}{\tikz[remember picture,baseline=-0.5ex]{\node[minimum
      height=2em,fill=psorange!50] (concrete-patt-value)
      {$v$}}}{\tikz[remember picture,baseline=-0.5ex]{\node[minimum
      height=2em,fill=psgreen!50] (concrete-patt-store)
      {$\sigma$}}}{\tikz[remember picture,baseline=-0.5ex]{\node[minimum
      height=2em,fill=psblue!50] (concrete-patt-envs)
      {$\esequence{\rho}$}}} \qquad
  \amatch{\tikz[remember picture,baseline=-0.5ex]{\node[minimum
      height=2em,fill=psred!50] (abs-patt)
      {$p$}}}{\tikz[remember picture,baseline=-0.5ex]{\node[minimum
      height=2em,fill=psorange!50] (abs-patt-value)
      {$\abst{\mathit{vs}}$}}}{\tikz[remember picture,baseline=-0.5ex]{\node[minimum
      height=2em,fill=psgreen!50] (abs-patt-store)
      {$\abst{\sigma}$}}}{\tikz[remember picture,baseline=-0.5ex]{\node[minimum
      height=2em,fill=psblue!50] (abs-patt-envs)
      {$\abst{\Rho}$}}}
\end{gather*}
\vspace{6em}
  \begin{gather*}
  \matchall{\tikz[remember picture,baseline=-0.5ex]{\node[minimum
      height=2em,fill=psred!50] (concrete-star-patt)
      {$\esequence{{\star}p}$}}}{\tikz[remember picture,baseline=-0.5ex]{\node[minimum
      height=2em,fill=psorange!50] (concrete-star-patt-val)
      {$\esequence{v}$}}}{\tikz[remember picture,baseline=-0.5ex]{\node[minimum
      height=2em,fill=psgreen!50] (concrete-star-patt-store)
      {$\sigma$}}}{\tikz[remember picture,baseline=-0.5ex]{\node[minimum
      height=2em,fill=psyellow] (concrete-star-backtrack)
      {$\mathbb{V}$}}}{}{\tikz[remember picture,baseline=-0.5ex]{\node[minimum
      height=2em,fill=psblue!50] (concrete-star-patt-envs)
      {$\esequence{\rho}$}}} \qquad\qquad
  \amatchall{\tikz[remember picture,baseline=-0.5ex]{\node[minimum
      height=2em,fill=psred!50] (abs-star-patt)
      {$\esequence{{\star}p}$}}}{\tikz[remember picture,baseline=-0.5ex]{\node[minimum
      height=2em,fill=psorange!50] (abs-star-patt-val)
      {$\abst{\mathit{vs}}$}}}{\tikz[remember picture,baseline=-0.5ex]{\node[minimum
      height=2em,fill=psgreen!50] (abs-star-patt-store)
      {$\abst{\sigma}$}}}{\tikz[remember picture,baseline=-0.5ex]{\node[minimum
      height=2em,fill=psorange!50] (abs-star-patt-length)
      {$[l;u]$}}}{\tikz[remember picture,baseline=-0.5ex]{\node[minimum
      height=2em,fill=psblue!50] (abs-star-patt-envs)
      {$\abst{\Rho}$}}}
\end{gather*}
\vspace{2em}
  \begin{tikzpicture}[remember picture,overlay]
  \draw[->,psred] (abs-patt) -- ++(0,1.5em) -| (concrete-patt) node[above,pos=.43] {\small same pattern};
  \draw[->,psgreen!120] (abs-patt-store) -- ++(0,-1.5em) -| (concrete-patt-store) node[below,pos=.4] {\small abstracts store};
  \draw[->,psblue] (abs-patt-envs) -- ++(0,+1.7em) -| (concrete-patt-envs) node[above,pos=.2] {\small abstracts binding environment sequence};
  \draw[->,psorange] (abs-patt-value) -- ++(0,-1.7em) -| (concrete-patt-value) node[below,pos=.3] {\small abstracts input value};
  \draw[->,dashed,psblue] (abs-patt-envs) -- ++(2.1em,0) -| ++(0,-3em) -| (abs-patt-store) node[below,pos=.2] {\small refines abstract store};
  \draw[->,dashed,psblue] (abs-patt-envs) -- ++(0,-1.5em) -| (abs-patt-value) node[below,pos=.2] {\small refines abstract value};
\end{tikzpicture}
  \begin{tikzpicture}[remember picture,overlay]
  \draw[->,psred] (abs-star-patt) -- ++(0,1.5em) -| (concrete-star-patt) node[above,pos=.43] {\small same pattern};
  \draw[->,psgreen!120] (abs-star-patt-store) -- ++(0,-1.5em) -| (concrete-star-patt-store) node[below,pos=.43] {\small abstracts store};
  \draw[->,psblue] (abs-star-patt-envs) -- ++(0,+2.8em) -| (concrete-star-patt-envs) node[above,pos=.2] {\small abstracts binding environment sequence};
  \draw[->,psorange] (abs-star-patt-val) -- ++(0,-1.7em) -| (concrete-star-patt-val) node[below,pos=.27] {\small abstracts shape of input value sequence};
  \draw[->,psorange] (abs-star-patt-length) -- ++(0,-3.2em) -| (concrete-star-patt-val) node[below,pos=.27] {\small abstracts length of input value sequence};
  \draw[->,dashed,psblue] (abs-star-patt-envs) -- ++(0,1.7em) -| (abs-star-patt-store) node[above,pos=.2] {\small refines abstract store};
  \draw[->,dashed,psblue] (abs-star-patt-envs) -- ++(2.2em,0) -| ++(0,-3.4em) -| (abs-star-patt-val) node[below,pos=.2] {\small refines shape};
  \draw[->,dashed,psblue] (abs-star-patt-envs) -- ++(0,-1.7em) -| (abs-star-patt-length) node[below,pos=.04] {\small refines length};
\end{tikzpicture}
  \end{minipage}
  \caption{Relating abstract operational semantics (left) to the
    concrete operational semantics (right).}
  \label{fig:abspattrules}
\end{figure*}
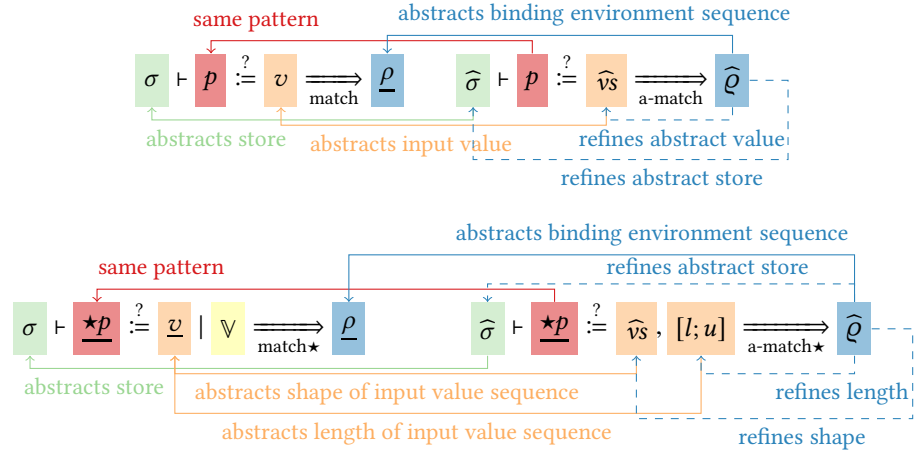

\paragraph{Computing Pattern Matching}
The judgements are presented in
\cref{fig:abspattrules} for both the concrete and abstract rules. Consider the
concrete (top-left) judgement: a value $v$ matches a pattern  $p$, given a store
$\sigma$, producing a sequence of binding environments $\esequence{\rho}$. The
binding environments form a sequence, since multiple concrete environments, say $\rho_1$ and $\rho_2$, can make $v$ match against $p$, i.e.,
$\satsem{v}{\rho_1}{p}$ and $\satsem{v}{\rho_2}{p}$. Backtracking using the
$\keyword{fail}$-expression, allows the programmer to explore a different
assignment from the sequence of environments, until no possible assignment is left.

For an ordinary pattern $p$ (top) the abstraction relation is direct: an abstract store  $\abst{\sigma}$ abstracts a concrete store $\sigma$ and a value shape $\abst{\mathit{vs}}$ abstracts a concrete value $v$.  The notable change is that the abstract semantics uses a set of abstract binding environments $\abst{\Rho} \subseteq \anamedset{Store} \times \anamedset{ValueShape} \times \anamedset{BindingEnv}_{\bot}$ that not only abstracts over the sequence of concrete binding environments $\esequence{\rho}$, but also, for each abstract binding environment stores the corresponding refinement of the input abstract store $\abst{\sigma}$ and the corresponding refinement of the matched value shape $\abst{\mathit{vs}}$ according to the matched pattern.

For sequences of set sub-patterns $\esequence{{\star}p}$, the sequence of
concrete values $\esequence{v}$ is abstracted by two components: the shape of
values $\abst{\mathit{vs}}$ and an interval approximating the length of the value sequence $[l;u]$. Both
of these values are refined as a result of the matching, which is captured by the abstract binding environment
$\abst{\Rho}$ (of the same type as for the simple patterns), since we treat the value refined as the abstract set containing the values
of the given shape and of given cardinality.
The concrete semantics of set sub-patterns also contains a backtracking state $\mathbb{V}$ which is
not used in the abstract semantics, because the abstraction of set elements is
coarse and we thus abstractly consider all possible subset assignments at the same time (joining instead of backtracking).

\paragraph{Operational Rules}
We will show how refinement is calculated by the abstract operational semantics by presenting some of key rules for abstract pattern matching.
Rascal also allows non-linear pattern matching against assigned store variables, and it is possible to use this information for refining the input store and abstract value.
In the \textsc{AP-V-U} rule we match the variable to the value shape and restrict the shape abstraction for the variable value to match the pattern.  The binding environment does not change as the name is already bound in the store.  In the \textsc{AP-V-F} rule, the matching fails ($\bot$), and then we learn that the value shape in the store should be refined to something that does not match.
\begin{gather*}
\inference[\textsc{AP-V-U}]{\abst{\sigma}(x) = (b, \abst{\mathit{vs}}') &
  \abst{\mathit{vs}}' \neq \bot_{\anamedset{VS}} \\
  \abst{\mathit{vs}}'' \in (\abst{\mathit{vs}} \abst{=} \abst{\mathit{vs}}')  &
  \abst{\sigma}' =  \abst{\sigma}[x \mapsto (\auxiliary{ff}, \abst{\mathit{vs}}'')]}{\amatchc{x}{\abst{\mathit{vs}}}{\abst{\sigma}}{}{(\abst{\sigma'}, \abst{\mathit{vs}}'', [])}{v}}
\end{gather*}
\begin{gather*}
\inference[\textsc{AP-V-F}]{\abst{\sigma}(x) = (b, \abst{\mathit{vs}}') &
  \abst{\mathit{vs}}' \neq \bot_{\anamedset{VS}} \\
  (\abst{\mathit{vs}}'',\abst{\mathit{vs}}''') \in (\abst{\mathit{vs}}
  \abst{\neq} \abst{\mathit{vs}}') & \abst{\sigma}' = \abst{\sigma}[x \mapsto (\auxiliary{ff}, \abst{\mathit{vs}}''')]}{\amatchc{x}{\abst{\mathit{vs}}}{\abst{\sigma}}{}{(\abst{\sigma}', \abst{\mathit{vs}}'', \bot)}{v}}
\end{gather*}
We also show the \textsc{AP-V-B} (abstract pattern-variable-bind) rule which simply binds the variable in the binding environment, assuming that it is possibly not assigned in the store (a free name).
\begin{gather*}
\inference[\textsc{AP-V-B}]{\abst{\sigma}(x) = (\auxiliary{tt},
  \abst{\mathit{vs}}')}{\amatchc{x}{\abst{\mathit{vs}}}{\abst{\sigma}}{}{(\abst{\sigma}[x
    \mapsto (\auxiliary{tt}, \bot_{\anamedset{VS}})], \abst{\mathit{vs}}, [x
    \mapsto \abst{\mathit{vs}}])}{v}}
\end{gather*}

If our matched abstract value possibly
contains the pattern constructor $k$ (\textsc{AP-C-S} rule: abstract pattern-constructor-success) we produce an abstract
value with $k$ containing the sub-values refined against constructor
sub-patterns:
\begin{gather*}
  \inference[\textsc{AP-C-S}]{\keyword{data} \; \mathit{at} = \dots \mid
    k(\esequence{t}) \mid \dots \\ (\keyword{success} \; k(\esequence{\abst{\mathit{vs}}'})) \in
    \aauxiliary{unfold}(\abst{\mathit{vs}}, \mathit{at})
\\  \amatch{p_1}{\abst{\mathit{vs}}'_1}{\abst{\sigma}}{\abst{\Rho}_1}
          \dots
          \amatch{p_{\mathrm{n}}}{\abst{\mathit{vs}}'_{\mathrm{n}}}{\abst{\sigma}}{\abst{\Rho}_{\mathrm{n}}}
          \\
          (\abst{\sigma}'_1, \abst{\mathit{vs}}'_1, \maybe{\abst{\rho}_1}) \in
            \abst{\Rho}_1
            \dots
            (\abst{\sigma}'_{\mathrm{n}}, \abst{\mathit{vs}}'_{\mathrm{n}}, \maybe{\abst{\rho}_{\mathrm{n}}}) \in
            \abst{\Rho}_{\mathrm{n}}
          }{\amatchc{k(\esequence{p})}{\abst{\mathit{vs}}}{\abst{\sigma}}{}{
(\bigsqcap_i \abst{\sigma}_i, k(\esequence{\abst{vs}'}), \aauxiliary{merge}(\esequence{\maybe{\abst{\rho}}}))}{cons}}
\end{gather*}
The total function \aauxiliary{merge} unifies assignments from two binding environments point-wise by names, taking the greatest lower bound of shapes to combine bindings for a name. It yields bottom for the entire result if at least one of the point-wise meets yields bottom (shapes for at least one name are not reconcilable).
Otherwise, we try to refine the matched value to exclude the
pattern constructor in the \textsc{AP-C-F} rules:

\begin{gather*}
  \inference[\textsc{AP-C-F1}]{\keyword{data} \; \mathit{at} = \dots \mid
    k(\esequence{t}) \mid \dots  \\ (\keyword{success} \; k'(\esequence{\abst{\mathit{vs}}'})) \in
\aauxiliary{unfold}(\abst{\mathit{vs}}, \mathit{at}) & k' \neq k }{\amatchc{k(\esequence{p})}{\abst{\mathit{vs}}}{\abst{\sigma}}{}{(\abst{\sigma},
      \aauxiliary{exclude}(\abst{\mathit{vs}}, k), \bot)}{cons}}
\end{gather*}

\begin{gather*}
  \inference[\textsc{AP-C-F2}]{\keyword{data} \; \mathit{at} = \dots \mid
    k(\esequence{t}) \mid \dots  &
\keyword{error} \in
\aauxiliary{unfold}(\abst{\mathit{vs}}, \mathit{at})}{\amatchc{k(\esequence{p})}{\abst{\mathit{vs}}}{\abst{\sigma}}{}{(\abst{\sigma},
      \aauxiliary{exclude}(\abst{\mathit{vs}}, k), \bot)}{cons}}
\end{gather*}

\noindent{}For set patterns, the refinement happens by pattern matching set sub-patterns.
  \begin{gather*}
   \inference[\textsc{AP-S-S}]{
 \keyword{success} \; \{\abst{\mathit{vs}}'\}_{[l;u]} \in  \aauxiliary{unfold}(\abst{\mathit{vs}},
 \keyword{set}\ttuple{\keyword{value}}) \\ \amatchall{\esequence{{{\star}p}}}{\abst{\mathit{vs}}}{\abst{\sigma}}{[l;u]}{\abst{\Rho}}
       }{\amatchc{\{\esequence{{{\star}p}}\}}{\abst{\mathit{vs}}}{\abst{\sigma}}{}{\abst{\Rho}}{set}}
  \end{gather*}

For example, when it is possible that the abstracted value sequence
($\abst{\mathit{vs}}, [l;u]$) is empty ($l = 0$) and patterned matched against
an empty set sub-pattern sequence, we can refine the result to be the empty
abstract set $\{\bot\}_0$ (rule \textsc{APL-E-B}).
\begin{gather*}
  \inference[\textsc{APL-E-B}]{l \leq u & l = 0}{\amatchallc{\varepsilon}{\abst{\mathit{vs}}}{\abst{\sigma}}{[l;u]}{
      (\abst{\sigma}, \{\bot_{\anamedset{VS}}\}_0, \{[]\}) }{1}}
\end{gather*}

\noindent
A more complex example is the one where we try to pattern match a potentially
non-empty value sequence against a set sub-pattern sequence $p, \esequence{{\star}p'}$ starting with an
ordinary pattern (\textsc{APL-M-P}). Here we pattern match against $p$ and the
rest of the sequence $\esequence{{\star}p'}$ and combine the refined results of
these matches producing a refinement of the containing set value by combining
the refined shapes and increasing the refinement of the length by the
set sub-pattern sequence by one.
   \begin{gather*}
  \inference[\textsc{APL-M-P}]{
    l \leq u &
         u \neq 0 &
                    \amatch{p}{\abst{\mathit{vs}}}{\abst{\sigma}}{\abst{\Rho_R}'}
                 \\
    \amatchall{\esequence{{{\star}p}}}{\abst{\mathit{vs}}}{\abst{\sigma}}{[l-1;u-1]}{\abst{\Rho_R}''}
                   \\
                         (\abst{\sigma}', \abst{\mathit{vs}}', \abst{\Rho}') \in
                         \abst{\Rho_R}' &
                         (\abst{\sigma}'', \{\abst{\mathit{vs}}''\}_{[l'';u'']}, \abst{\Rho}'') \in
                         \abst{\Rho_R}'' \\
                         \abst{\Rho_R}''' = \left\{
                         {\begin{lgathered}
                         (\abst{\sigma}' \sqcap \abst{\sigma}'', 
                                        \{\abst{\mathit{vs}}'\sqcup
                                        \abst{\mathit{vs}}''\}_{[l''+1,u''+1]} , \\
                                        \aauxiliary{merge}(\abst{\Rho}',
                                        \abst{\Rho}''))
                         \end{lgathered}}
                                        \right\}
  }{\amatchallc{p,\esequence{{{\star}p}}}{\abst{\mathit{vs}}}{\abst{\sigma}}{[l;u]}{\abst{\Rho_R}'''}{1}}
\end{gather*}

\section{Abstract Semantic Rules}
\label{sec:abssemanticrules}
\Cref{fig:absvisitorrules,fig:absvisitorrules2} shows the formal rules for executing the
bottom-up visit-expression; we have omitted the collecting rules and some
error handling rules to avoid presenting unnecessary details.
We will further discuss the ideas behind the rules in a high-level fashion.

\paragraph{Executing visitors}
The evaluation rule for the $\keyword{visit}$-expression itself is mainly
concerned with evaluating the target expression $e$ to be traversed to a value, and
then using a separate traversal relation to rewrite the value recursively with
the sequence of cases $\esequence{\mathit{cs}}$. The main item to notice is how
it uses the value refined by the case patterns in case of failure
(\rulename{AE-Vt-F}), turning the result into successful execution (like in our
running example in \cref{sec:motivationandidea}).
\paragraph{Evaluating Cases}
During traversal, the target value will be rewritten with a sequence of cases.
The evaluation of a case sequence is straight-forward, iterating through the
possible cases, pattern matching against each pattern and executing the
corresponding expression when applicable. The main idea is that, when
the abstract value fails to match a pattern, the refined value is used to match
against the rest of the cases (\rulename{ACS-M-F}). This ensures that the order
of patterns influences the refinement, leading to a more precise abstract shape
that better matches the set of concrete shapes during execution.

\begin{figure*}[p]
  \centering
\fbox{
\begin{minipage}{.95\textwidth}
  \textbf{Expressions (General)}%
  \begin{center}
  {\small\renewcommand*{\arraystretch}{3}
    \begin{tabular}{ccc}
  $\inference[\rulename{AE-A}]{\collect{\aevalexprc{x =
        e}{\abst{\sigma}}{\abst{\mathit{Res}}}{asgn}}}{\aevalexpr{x =
      e}{\abst{\sigma}}{\abst{\mathit{Res}}}}$ &
  $\inference[\rulename{AE-Sq}]{\collect{\aevalexprc{e_1;e_2}{\abst{\sigma}}{\abst{\mathit{Res}}}{seq}}}{\aevalexpr{e_1; e_2}{\abst{\sigma}}{\abst{\mathit{Res}}}}$ &
    $\inference[\rulename{AE-C}]{\collect{\aevalexprc{k(\esequence{e})}{\abst{\sigma}}{\abst{\mathit{Res}}}{cons}}}{\aevalexpr{k(\esequence{e})}{\abst{\sigma}}{\abst{\mathit{Res}}}}$ \\
  $\inference[\rulename{AE-St}]{\collect{\aevalexprc{\{\esequence{e}\}}{\abst{\sigma}}{\abst{\mathit{Res}}}{set}}}{\aevalexpr{\{\esequence{e}\}}{\abst{\sigma}}{\abst{\mathit{Res}}}}$ &
                                                                                                                                                                                         $\inference[\rulename{AE-Fl}]{}{\aevalexpr{\keyword{fail}}{\abst{\sigma}}{[\keyword{fail} \mapsto \tuple{\cdot, \abst{\sigma}}]}}$ &
$\inference[\rulename{AES}]{\collect{\aevalexprstarc{\esequence{e}}{\abst{\sigma}}{\abst{\mathit{Res}}}{1}}}{\aevalexprstar{\esequence{e}}{\abst{\sigma}}{\abst{\mathit{Res}}}}$
    \end{tabular}}
  \end{center}

    \textbf{Assignment Expression}
    \vspace{-1.5em}
    \begin{center}
     {\small
  \begin{gather*}
    \inference[\rulename{AE-A-S}]{
      \keyword{local} \; t \; x \vee \keyword{global} \; t \; x &
      \aevalexpr{e}{\abst{\sigma}}{\abst{\mathit{Res}}} \\
      \tuple{\keyword{success}, \tuple{\abst{vs}, \abst{\sigma}'}} \in  \abst{\mathit{Res}} &
      \atyping{\abst{vs}}{t'} & \asubtyping{t'}{t}
    }{\aevalexprc{x = e}{\abst{\sigma}}{[\keyword{success} \mapsto \tuple{\abst{vs},
          \abst{\sigma}'[x \mapsto (\auxiliary{ff}, \abst{vs})]}]}{asgn}}
    \end{gather*}
    \begin{gather*}
    \inference[\rulename{AE-A-Er}]{
      \keyword{local} \; t \; x \vee \keyword{global} \; t \; x &
      \aevalexpr{e}{\abst{\sigma}}{\abst{\mathit{Res}}} \\
      \tuple{\keyword{success}, \tuple{\abst{vs}, \abst{\sigma}'}} \in  \abst{\mathit{Res}} &
      \atyping{\abst{vs}}{t'} & \anotsubtyping{t'}{t}
    }{\aevalexprc{x = e}{\abst{\sigma}}{[\keyword{error} \mapsto \tuple{\cdot, \abst{\sigma}'}]}{asgn}}
    \qquad
        \inference[\rulename{AE-A-Ex}]{
      \aevalexpr{e}{\abst{\sigma}}{\abst{\mathit{Res}}} &
      \tuple{\mathit{exres}, \tuple{\abst{\mathit{resv}}, \abst{\sigma}'}} \in
       \abst{\mathit{Res}}}{\aevalexprc{x =
        e}{\abst{\sigma}}{[\mathit{exres} \mapsto \tuple{\abst{\mathit{resv}}, \abst{\sigma}'}]}{asgn}}
  \end{gather*}}
    \end{center}
        \textbf{Sequencing Expression}
    \vspace{-1em}
    {\small
    \begin{center}
    \begin{gather*}
      \inference[\rulename{AE-Sq-S}]{\aevalexprstar{e_1,e_2}{\abst{\sigma}}{\abst{\mathit{Res}{\star}}}
        \\ \tuple{\keyword{success}, \tuple{(\abst{\mathit{vs}}_1, \abst{\mathit{vs}}_2),
            \abst{\sigma}'}} \in  \abst{\mathit{Res}{\star}}}{\aevalexprc{e_1;e_2}{\abst{\sigma}}{[\keyword{success}
      \mapsto \tuple{\abst{\mathit{vs}}_2, \abst{\sigma}'}]}{seq}} \qquad
            \inference[\rulename{AE-Sq-Ex}]{\aevalexprstar{e_1,e_2}{\abst{\sigma}}{\abst{\mathit{Res}{\star}}}
              \\ \tuple{\mathit{exres},
                \tuple{\abst{\mathit{resv}}, \abst{\sigma}'}} \in
              \abst{\mathit{Res}{\star}}}{\aevalexprc{e_1;e_2}{\abst{\sigma}}{[\mathit{exres}
                \mapsto \tuple{\abst{\mathit{resv}}, \abst{\sigma}'}]}{seq}}
    \end{gather*}
    \end{center}}
      \textbf{Constructor Expression}
  \vspace{-1em}
  \begin{center}
  {\small
  \begin{gather*}
    \inference[\rulename{AE-C-S}]{\keyword{data} \; \mathit{at} = \dots |
      k(\esequence{t}) | \dots &
      \aevalexprstar{\esequence{e}}{\abst{\sigma}}{\abst{\mathit{Res}{\star}}} \\
      \tuple{\keyword{success}, \tuple{\esequence{\abst{\mathit{vs}}},
        \abst{\sigma}'}} \in
    \abst{\mathit{Res}{\star}} & \esequence{\atyping{\abst{\mathit{vs}}}{t'}} &
    \esequence{\asubtyping{t'}{t}}}{\aevalexpr{k(\esequence{e})}{\abst{\sigma}}{[\keyword{success}
      \mapsto \tuple{k(\esequence{\abst{\mathit{vs}}}), \abst{\sigma}'}}]}
  \end{gather*}
  \begin{gather*}
  \inference[\rulename{AE-C-Er}]{\keyword{data} \; \mathit{at} = \dots |
      k(\esequence{t}) | \dots &
      \aevalexprstar{\esequence{e}}{\abst{\sigma}}{\abst{\mathit{Res}{\star}}} \\
      \tuple{\keyword{success}, \tuple{\esequence{\abst{\mathit{vs}}}, \abst{\sigma}'}} \in
       \abst{\mathit{Res}{\star}} & \esequence{\atyping{\abst{\mathit{vs}}}{t'}} &
    \exists i.
    \anotsubtyping{t'_i}{t_i}}{\aevalexpr{k(\esequence{e})}{\abst{\sigma}}{[\keyword{error} \mapsto
      \tuple{\cdot, \abst{\sigma}'}]}} \qquad
\inference[\rulename{AE-C-Ex}]{\aevalexprstar{\esequence{e}}{\abst{\sigma}}{\abst{\mathit{Res}{\star}}} &
  \tuple{\mathit{exres}, \tuple{\abst{\mathit{resv}}, \abst{\sigma}'}} \in
   \abst{\mathit{Res}{\star}}}{\aevalexpr{k(\esequence{e})}{\abst{\sigma}}{
      [\mathit{exres} \mapsto \tuple{\abst{\mathit{resv}}, \abst{\sigma}'}]}}
  \end{gather*}}
  \end{center}

    \textbf{Set Expression}
  \vspace{-1em}
  \begin{center}
  {\small
  \begin{gather*}
    \inference[\rulename{AE-St-S}]{\aevalexprstar{\esequence{e}}{\abst{\sigma}}{\abst{\mathit{Res}{\star}}}
      \\ \tuple{\keyword{success}, \tuple{\esequence{\mathit{vs}}, \abst{\sigma}'}} \in
       \abst{\mathit{Res}{\star}}}{\aevalexprc{\{\esequence{e}\}}{\abst{\sigma}}{[\keyword{success}
        \mapsto
        \tuple{\{\bigsqcup_i \abst{\mathit{vs}_i}\}_{[0;|\esequence{\mathit{vs}}| ]}, \abst{\sigma}'}]}{set}} \qquad
       \inference[\rulename{AE-St-Ex}]{\aevalexprstar{\esequence{e}}{\abst{\sigma}}{\abst{\mathit{Res}{\star}}}
         \\ \tuple{\mathit{exres}, \tuple{\abst{\mathit{resv}},
             \abst{\sigma}'}} \in
         \abst{\mathit{Res}{\star}}}{\aevalexprc{\{\esequence{e}\}}{\abst{\sigma}}{[\mathit{exres}
           \mapsto \tuple{\abst{\mathit{resv}}, \abst{\sigma}'}]}{set}}
  \end{gather*}}
  \end{center}
\textbf{Expression Sequences}

\begin{center}
  {\small
  \begin{tabular}{cc}
     $\inference[\rulename{AES-Em}]{}{\aevalexprstarc{\varepsilon}{\abst{\sigma}}{[\keyword{success} \mapsto \tuple{\varepsilon, \abst{\sigma}}]}{1}}$ &
    $\inference[\rulename{AES-Mr}]{\aevalexpr{e}{\abst{\sigma}}{\abst{\mathit{Res}}} & \tuple{\keyword{success}, \tuple{\abst{vs}, \abst{\sigma}''}} \in  \abst{\mathit{Res}} \\
    \aevalexprstar{\esequence{e'}}{\abst{\sigma}''}{\abst{\mathit{Res}{\star}}'}  & \tuple{\keyword{success}, \tuple{\esequence{\abst{vs}'}, \abst{\sigma}'}} \in  \abst{\mathit{Res}{\star}}'
                                                                                    }{\aevalexprstarc{e,\esequence{e'}}{\abst{\sigma}}{[\keyword{success} \mapsto \tuple{\tuple{\abst{vs}, \esequence{\abst{vs}'}}, \abst{\sigma}'}]}{1}}$ \\[3em]
    $\inference[\rulename{AES-Ex}]{\aevalexpr{e}{\abst{\sigma}}{\abst{\mathit{Res}}} & \tuple{\mathit{exres}, \tuple{\abst{\mathit{resv}}, \abst{\sigma}'}} \in  \abst{\mathit{Res}}}{\aevalexprstarc{e,\esequence{e'}}{\abst{\sigma}}{[\mathit{exres} \mapsto \tuple{\abst{\mathit{resv}}, \abst{\sigma}'}]}{1}}$
    & $\inference[\rulename{AES-Ex}]{\aevalexpr{e}{\abst{\sigma}}{\abst{\mathit{Res}}} & \tuple{\keyword{success}, \tuple{\abst{vs}, \abst{\sigma}''}} \in  \abst{\mathit{Res}} \\
    \aevalexprstar{\esequence{e'}}{\abst{\sigma}''}{\abst{\mathit{Res}{\star}}'}  & \tuple{\mathit{exres}, \tuple{\abst{\mathit{resv}}, \abst{\sigma}'}} \in  \abst{\mathit{Res}{\star}}'
                                                                                    }{\aevalexprstarc{e,\esequence{e'}}{\abst{\sigma}}{[\mathit{exres} \mapsto \tuple{\abst{\mathit{resv}}, \abst{\sigma}'}]}{1}}$
      \end{tabular}}
\end{center}
    \end{minipage}
}
  \caption{Abstract Operational Semantics Rules for Basic Expressions}
  \label{fig:absrascalrules}
\end{figure*}

\begin{figure*}[htbp]
  \centering
\fbox{
\begin{minipage}{.95\textwidth}
  \textbf{Visit Expression}%
  \vspace{-1em}
  \begin{center}
  {\small
\begin{gather*}
  \inference[\rulename{AE-Vt-S}]{
\aevalexpr{e}{\abst{\sigma}}{\abst{\mathit{Res}}}
& \tuple{\keyword{success},  \tuple{\abst{\mathit{vs}}, \abst{\sigma}''}} \in
     \abst{\mathit{Res}} \\
    \aevalbuvisit{\esequence{\mathit{cs}}}{\abst{\mathit{vs}}}{\abst{\sigma}''}{\abst{\mathit{Res}}'}
    & \tuple{\keyword{success}, \tuple{\abst{\mathit{vs}}', \abst{\sigma}'}} \in
     \abst{\mathit{Res}}'
  }{    \aevalexprc{\keyword{visit} \; e \;
        \esequence{\mathit{cs}}}{\abst{\sigma}}{[\keyword{success} \mapsto
        \tuple{\abst{\mathit{vs}}', \abst{\sigma}'}]}{visit}}
    \qquad
    \inference[\rulename{AE-Vt-F}]{
\aevalexpr{e}{\abst{\sigma}}{\abst{\mathit{Res}}}
& \tuple{\keyword{success}, \tuple{\abst{\mathit{vs}}, \abst{\sigma}''}} \in
     \abst{\mathit{Res}} \\
    \aevalbuvisit{\esequence{\mathit{cs}}}{\abst{\mathit{vs}}}{\abst{\sigma}''}{\abst{\mathit{Res}}'}
    & \tuple{\keyword{fail}, \tuple{\abst{\mathit{vs}}', \abst{\sigma}'}} \in
    \abst{\mathit{Res}}'
  }{    \aevalexprc{\keyword{visit} \; e \;
        \esequence{\mathit{cs}}}{\abst{\sigma}}{[\keyword{success} \mapsto
        \tuple{\abst{\mathit{vs}}', \abst{\sigma}'}]}{visit}}
 \end{gather*}
\begin{gather*}
      \inference[\rulename{AE-Vt-Ex1}]{
\aevalexpr{e}{\abst{\sigma}}{\abst{\mathit{Res}}}
& \tuple{\mathit{exres}, \tuple{\abst{\mathit{resv}}, \abst{\sigma}'}}
\in
    \abst{\mathit{Res}}   }{    \aevalexprc{\keyword{visit} \; e \;
        \esequence{\mathit{cs}}}{\abst{\sigma}}{[\mathit{exres} \mapsto
        \tuple{\abst{\mathit{resv}}, \abst{\sigma}'}]}{visit}}
    \qquad
     \inference[\rulename{AE-Vt-Ex2}]{
\aevalexpr{e}{\abst{\sigma}}{\abst{\mathit{Res}}}
& \tuple{\keyword{success}, \tuple{\abst{\mathit{vs}}, \abst{\sigma}''}} \in
    \abst{\mathit{Res}} \\
    \aevalbuvisit{\esequence{\mathit{cs}}}{\abst{\mathit{vs}}}{\abst{\sigma}''}{\abst{\mathit{Res}}'}
    & \tuple{\keyword{error}, \tuple{\abst{\mathit{resv}}, \abst{\sigma}'}} \in
    \abst{\mathit{Res}}'
  }{    \aevalexprc{\keyword{visit} \; e \;
        \esequence{\mathit{cs}}}{\abst{\sigma}}{[\keyword{error} \mapsto \tuple{
          \abst{\mathit{resv}}, \abst{\sigma}'}]}{visit}}
  \end{gather*}}
  \end{center}

  \textbf{Bottom-up Traversal of Single Value}
  {\small
  \begin{gather*}
  \inference[\rulename{ABU-S}]{ \tuple{\abst{\mathit{vs}}'',
      \abst{\mathit{cvs}}} \in
    \aauxiliary{children}(\abst{\mathit{vs}}) &
    \aevalbuvisitstar{\esequence{\mathit{cs}}}{\abst{\mathit{cvs}}}{\abst{\sigma}}{\abst{\mathit{Res}{\star}}}
    &
    \tuple{\keyword{success}, \tuple{\abst{\mathit{cvs}}',
      \abst{\sigma}'}} \in \abst{\mathit{Res}{\star}} \\
  \areconstruct{\abst{\mathit{vs}}''}{\abst{\mathit{cvs}}'}{\abst{\mathit{RCRes}}}
    &
    \tuple{\keyword{success}, \abst{\mathit{vs}}'} \in \abst{\mathit{RCRes}} &
    \aevalcases{\esequence{\mathit{cs}}}{\abst{\mathit{vs}}'}{\abst{\sigma}'}{\abst{\mathit{Res}}'}
  }{\aevalbuvisitc{\esequence{\mathit{cs}}}{\abst{\mathit{vs}}}{\abst{\sigma}}{\abst{\mathit{Res}}'}{go}}
  \end{gather*}
  \begin{gather*}
      \inference[\rulename{ABU-F}]{ \tuple{\abst{\mathit{vs}}'',
      \abst{\mathit{cvs}}} \in
    \aauxiliary{children}(\abst{\mathit{vs}}) &
    \aevalbuvisitstar{\esequence{\mathit{cs}}}{\abst{\mathit{cvs}}}{\abst{\sigma}}{\abst{\mathit{Res}{\star}}}
    \\
    \tuple{\keyword{fail}, \tuple{\abst{\mathit{cvs}}',
        \abst{\sigma}'}} \in
    \abst{\mathit{Res}{\star}} &
    \areconstruct{\abst{\mathit{vs}}''}{\abst{\mathit{cvs}}'}{[\keyword{success}
      \mapsto \abst{\mathit{vs}}']} &
    \aevalcases{\esequence{\mathit{cs}}}{\abst{\mathit{vs}}'}{\abst{\sigma}'}{\abst{\mathit{Res}}'}
  }{\aevalbuvisitc{\esequence{\mathit{cs}}}{\abst{\mathit{vs}}}{\abst{\sigma}}{\abst{\mathit{Res}}'}{go}}
\end{gather*}}

\textbf{Bottom-up Traversal of Children}
\vspace{-1.2em}
{\small
  \begin{gather*}
    \inference[\rulename{ABUC-E}]{}{\aevalbuvisitstarc{\esequence{\mathit{cs}}}{\varepsilon}{\abst{\sigma}}{[\keyword{fail} \mapsto
        \tuple{\varepsilon,\abst{\sigma}}]}{go}}
    \quad
    \inference[\rulename{ABUC-M}]{\aevalbuvisit{\esequence{\mathit{cs}}}{\abst{\mathit{vs}}}{\abst{\sigma}}{\abst{\mathit{Res}}}
      & \tuple{\abst{\mathit{vfres}}, \tuple{\abst{\mathit{vs}}'', \abst{\sigma}''}} \in \abst{\mathit{Res}} \\
      \aevalbuvisitstar{\esequence{\mathit{cs}}}{\esequence{\abst{\mathit{vs}}'}}{\abst{\sigma}''}{\abst{\mathit{Res}{\star}}'}
      & \tuple{\abst{\mathit{vfres}}', \tuple{\esequence{\abst{\mathit{vs}}'''},
          \abst{\sigma}'}} \in \abst{\mathit{Res}{\star}}' \\
      \abst{\mathit{Res}}'' = \aauxiliary{vcombine}(\abst{\mathit{vfres}},
        \abst{\mathit{vs}}'', \abst{\mathit{vfres}}',
        \esequence{\abst{\mathit{vs}}'''}, \abst{\sigma}')
    }{\aevalbuvisitstarc{\esequence{\mathit{cs}}}{\abst{\mathit{vs}},
        \esequence{\abst{\mathit{vs}}'}}{\abst{\sigma}}{\abst{\mathit{Res}}''}{go}}
  \end{gather*}
  \begin{gather*}
    \inference[\rulename{ABUS-E}]{}{\aevalbuvisitstarc{\esequence{\mathit{cs}}}{\tuple{\abst{\mathit{vs}},[0;u]}}{\abst{\sigma}}{[\keyword{fail} \mapsto
        \tuple{\tuple{\bot,0},\abst{\sigma}}]}{go}}
    \\
    \inference[\rulename{ABUS-M}]{u > 0 & \aevalbuvisit{\esequence{\mathit{cs}}}{\abst{\mathit{vs}}}{\abst{\sigma}}{\abst{\mathit{Res}}}
      & \tuple{\abst{\mathit{vfres}}, \tuple{\abst{\mathit{vs}}'', \abst{\sigma}''}} \in \abst{\mathit{Res}} &
      \aevalbuvisitstar{\esequence{\mathit{cs}}}{\tuple{\abst{\mathit{vs}}, [l-1;u-1]}}{\abst{\sigma}''}{\abst{\mathit{Res}{\star}}'}
      \\ \tuple{\abst{\mathit{vfres}}', \tuple{\tuple{\abst{\mathit{vs}}''',[l';u']},
          \abst{\sigma}'}} \in \abst{\mathit{Res}{\star}}' &
      \abst{\mathit{Res}}'' = \aauxiliary{vcombine}(\abst{\mathit{vfres}},
        \abst{\mathit{vs}}'', \abst{\mathit{vfres}}', \tuple{\abst{\mathit{vs}}''',[l';u']},\abst{\sigma}')
      }{\aevalbuvisitstarc{\esequence{\mathit{cs}}}{\tuple{\abst{\mathit{vs}}, [l;u]}}{\abst{\sigma}}{\abst{\mathit{Res}}''}{go}}
  \end{gather*}}
  \end{minipage}
}
\vspace{-2mm}
  \caption{Selected Abstract Operational Semantics Rules for Traversal}
  \label{fig:absvisitorrules}
\end{figure*}

\begin{figure*}
  \centering
\fbox{
\begin{minipage}{.95\textwidth}
\textbf{Case Sequence}
\vspace{-1.4em}
{\small\begin{gather*}
    \inference[\rulename{ACS-E}]{}{\aevalcasesc{\varepsilon}{\abst{\mathit{vs}}}{\abst{\sigma}}{[\keyword{fail}
        \mapsto \tuple{\abst{\mathit{vs}}, \abst{\sigma}}]}{go}} \quad
    \inference[\rulename{ACS-M-O}]{\amatch{p}{\abst{\mathit{vs}}}{\abst{\sigma}}{\abst{\Rho}}
      & \tuple{\abst{\mathit{vs'}}, \abst{\sigma}', \maybe{\abst{\rho}}} \in
      \abst{\Rho} \\
      \aevalcase{\maybe{\abst{\rho}}}{e}{\abst{\sigma}'}{\abst{\mathit{Res}}} &
      \tuple{\mathit{rest}, \tuple{\abst{\mathit{resv}}, \abst{\sigma}''}} \in
      \abst{\mathit{Res}} & \mathit{rest} \neq
      \keyword{fail}}{\aevalcasesc{\keyword{case} \; p => e,
        \esequence{\mathit{cs}}}{\abst{\mathit{vs}}}{\abst{\sigma}}{[\mathit{rest}
        \mapsto \tuple{\abst{\mathit{resv}}, \abst{\sigma}''}]}{go}}
  \end{gather*}
  \begin{gather*}
    \inference[\rulename{ACS-M-F}]{
       \amatch{p}{\abst{\mathit{vs}}}{\abst{\sigma}}{\abst{\Rho}}
      & \tuple{\abst{\mathit{vs'}}, \abst{\sigma}', \maybe{\abst{\rho}}} \in
      \abst{\Rho} \\
      \aevalcase{\maybe{\abst{\rho}}}{e}{\abst{\sigma}'}{\abst{\mathit{Res}}} &
      \tuple{\keyword{fail}, \tuple{\abst{\mathit{resv}}, \abst{\sigma}''}} \in
      \abst{\mathit{Res}} &
 \aevalcasesc{\esequence{\mathit{cs}}}{\abst{\mathit{vs}}'}{\abst{\sigma}'}{\abst{\mathit{Res}}'}{go}
    }{
      \aevalcasesc{\keyword{case} \; p => e,
        \esequence{\mathit{cs}}}{\abst{\mathit{vs}}}{\abst{\sigma}}{\abst{\mathit{Res}}'}{go}
    }
  \end{gather*}
}

\textbf{Case}
\vspace{-1.1em}
{\small\begin{gather*}
    \inference[\rulename{AC-E}]{}{\aevalcasec{\bot}{e}{\abst{\sigma}}{[\keyword{fail}
        \mapsto \tuple{\cdot, \abst{\sigma}}]}{go}} \quad
    \inference[\rulename{AC-M-O}]{\aevalexpr{\abst{\sigma} \;
        \abst{\rho}}{e}{\abst{\mathit{Res}}} & \tuple{\mathit{rest}, \tuple{\abst{\mathit{resv}}, \abst{\sigma}''}} \in
      \abst{\mathit{Res}} & \mathit{rest} \neq
      \keyword{fail}}{\aevalcasec{\abst{\rho}}{\abst{\mathit{vs}}}{\abst{\sigma}}{[\mathit{rest}
        \mapsto \tuple{\abst{\mathit{resv}}, \abst{\sigma}''}]}{go}}
  \end{gather*}
  \begin{gather*}
        \inference[\rulename{AC-M-F}]{\aevalexpr{\abst{\sigma} \;
        \abst{\rho}}{e}{\abst{\mathit{Res}}} & \tuple{\keyword{fail}, \tuple{\abst{\mathit{resv}}, \abst{\sigma}''}} \in
      \abst{\mathit{Res}}}{\aevalcasec{\abst{\rho}}{\abst{\mathit{vs}}}{\abst{\sigma}}{[\keyword{fail}
        \mapsto \tuple{\abst{\mathit{resv}}, \abst{\sigma}}]}{go}}
  \end{gather*}
}
\end{minipage}
}
  \caption{Selected Abstract Operational Semantic Rules for Traversal (Cont.)}
  \label{fig:absvisitorrules2}
\end{figure*}

\begin{acks}
  We would like to thank Paul Klint, Tijs van der Storm, Jurgen Vinju and Davy
  Landman for discussions on Rascal and its semantics. We would further like to thank
 Rasmus M{\o}gelberg and Jan Midtgaard for discussions on correctness of our
 recursive shape abstractions. We would like the anonymous reviewers for their
 comments, especially the one who presented us the link between our style of
 abstract interpretation and verification techniques in program transformation.

  This material is based upon work supported by the
  \grantsponsor{GS501100004836}{Danish Council for Independent
    Research}{http://dx.doi.org/10.13039/501100004836} under Grant
  No.~\grantnum{GS501100004836}{0602-02327B} and
  \grantsponsor{GS100012774}{Innovation Fund Denmark}{http://dx.doi.org/10.13039/100012774} under Grant
  No.~\grantnum{GS100012774}{7039-00072B}.  Any opinions, findings, and
  conclusions or recommendations expressed in this material are those
  of the author and do not necessarily reflect the views of the
  funding agencies.
\end{acks}

\bibliography{paper}

\end{document}